\documentstyle[12pt,epsfig]{article}
\begin{document}
\thispagestyle{empty}
\def\ltap{\ \raisebox{-.4ex}{\rlap{$\sim$}} \raisebox{.4ex}{$<$}\ }
\def\gtap{\ \raisebox{-.4ex}{\rlap{$\sim$}} \raisebox{.4ex}{$>$}\ }
\setcounter{page}0
% \preprint{\vbox{\baselineskip 10pt{
\rightline{Ref. SISSA 28/99/EP}
\rightline{March 1999}
\rightline{hep -- ph/9903424}
\vskip 0.6cm
% ~\vfill
\begin{center}
{\bf
Enhancing Mechanisms of Neutrino Transitions in a Medium of \\ 
Nonperiodic Constant - Density Layers and in the Earth \\
} 

\vspace{0.3cm} 
M. V. Chizhov

\vspace{0.3cm}
{\em
Centre for Space Research and Technologies, Faculty of Physics,\\
University of Sofia, 1164
Sofia, Bulgaria\\
E-mail: $mih@phys.uni$-$sofia.bg$
}

\vspace{0.4cm}
S. T. Petcov \footnote{Also at: Institute of Nuclear Research and
Nuclear Energy, Bulgarian Academy of Sciences, 1784 Sofia, Bulgaria}

\vspace{0.3cm}   
{\em
Scuola Internazionale Superiore di Studi Avanzati, I-34014 Trieste,
Italy, and\\
Istituto Nazionale di Fizica Nucleare, Sezione di Trieste, I-34014
Trieste, Italy
}
\end{center}
% \vfill
\vskip 0.4cm
\begin{abstract}
The enhancement of the transitions
$\nu_{\mu}~(\nu_e) \rightarrow \nu_e~(\nu_{\mu;\tau})$,
$\nu_{2} \rightarrow \nu_e$,
$\bar{\nu}_{\mu} \rightarrow \bar{\nu}_s$, 
$\nu_e \rightarrow \nu_s$, etc. of neutrinos 
passing through
a medium of nonperiodic density distribution, 
consisting of i) two layers of different constant 
densities (e.g., Earth core and mantle),
or ii) three layers of constant density 
with the first and the third
layers having identical densities and widths 
which differ from those of the second layer
(e.g., mantle-core-mantle in the Earth), is studied.  
We find that, 
in addition to the known local maxima 
associated with the MSW effect and the 
neutrino oscillation length resonance (NOLR), 
the relevant two-neutrino 
transition probabilities have 
new resonance-like absolute maxima. 
The latter correspond to a new effect 
of {\it total} neutrino conversion and are 
absolute maxima in any 
independent variable: the neutrino energy,
the layers widths, etc.  
We find the complete set of such maxima.
We prove that the NOLR and  
the new absolute maxima 
are caused by a maximal constructive 
interference between the amplitudes 
of the neutrino transitions in the different 
density layers. 
We show that the strong resonance-like 
enhancement of the  transitions
in the Earth of the 
Earth-core-crossing solar 
and atmospheric neutrinos
is due to the new effect of total 
neutrino conversion. 

\end{abstract}

\vfill
\newpage
\section{Introduction} \indent 
  It was pointed out in \cite{SP1} that 
the $\nu_2 \rightarrow \nu_{e}$, 
$\nu_{\mu} \rightarrow \nu_{e}$
($\nu_e \rightarrow \nu_{\mu(\tau)}$) 
oscillations of solar and atmospheric neutrinos 
in the Earth, caused by
neutrino mixing (with nonzero mass neutrinos) in vacuum
\footnote{The $\nu_2 \rightarrow \nu_{e}$
transition probability accounts, as is well-known, for the Earth effect
in the solar neutrino survival probability
in the case of the MSW two-neutrino
$\nu_e \rightarrow \nu_{\mu(\tau)}$ and
$\nu_e \rightarrow \nu_{s}$ transition solutions of the
solar neutrino problem, $\nu_{s}$ being a sterile neutrino.},
can be strongly amplified by a
new type of resonance-like mechanism
which differs from the MSW one and
takes place when the neutrinos 
traverse the Earth core on the
way to the detector 
(see also \cite{s5398,SPnu98,SPNewE98}).
As numerical calculations have shown, 
at small mixing angles ($\sin^22\theta \ltap 0.05$),
the maxima in the neutrino energy $E$ 
of the corresponding transition probabilities,
$P(\nu_2 \rightarrow \nu_{e}) \equiv P_{e2}$ and
$P(\nu_{\mu} \rightarrow \nu_{e})$
($P(\nu_e \rightarrow \nu_{\mu(\tau)})$),
caused by the new enhancing mechanism, 
are absolute maxima and dominate in  $P_{e2}$ and
$P(\nu_{\mu} \rightarrow \nu_{e})$
($P(\nu_e \rightarrow \nu_{\mu(\tau)})$):
the values of the probabilities at these maxima
in the simplest case of two-neutrino mixing
are considerably larger -
by a factor of $\sim (2.5 - 4.0)$ ($\sim (3.0 - 7.0)$),
than the values of $P_{e2}$ and
$P(\nu_{\mu} \rightarrow \nu_{e}) = 
P(\nu_e \rightarrow \nu_{\mu(\tau)})$
at the local maxima
associated with the MSW effect
taking place in the Earth core (mantle). 
The effect of the new enhancement 
is equally dramatic at large mixing angles. 
The enhancement is present and dominates 
also in the $\nu_2 \rightarrow \nu_{e}$ transitions 
in the case of $\nu_e - \nu_{s}$ mixing and 
in the $\nu_e \rightarrow \nu_{s}$ and 
$\bar{\nu}_{\mu} \rightarrow \bar{\nu}_{s}$ 
transitions at small mixing angles 
\cite{SP1,s5398,SPnu98,SPNewE98}.
Even at small mixing angles 
the enhancement 
is relatively 
wide \cite{SP1,s5398,SPNewE98} 
in the Nadir angle \footnote{The Nadir angle
determines uniquely the neutrino trajectory through
the Earth.}, $h$, and in the neutrino energy 
and leads to important observable effects.
It also exhibits rather strong energy dependence.

  The new resonance-like enhancement
of the probability $P_{e2}$ has 
important implications 
for the tests of the MSW
$\nu_e \rightarrow \nu_{\mu (\tau)}$
and $\nu_e \rightarrow \nu_{s}$
transition solutions of the solar neutrino problem
\cite{SP1,SPnu98,Art2,Art3,Art1,MPSNODN00,MPDNCoreSK00}. 
% (see also \cite{Art1}).   
For values of
$\Delta m^2$ % \cong (4.0 - 8.0) \times 10^{-6}~{\rm eV^2}$
from the small mixing angle (SMA) MSW solution region
and the geographical latitudes
at which the Super-Kamiokande, SNO and ICARUS
detectors are located,
the enhancement takes place
in the $\nu_e \rightarrow \nu_{\mu (\tau)}$ 
case for values of the
$^{8}$B neutrino energy
lying in the interval
$\sim (5 - 12)~$MeV to which
these detectors are sensitive.
Accordingly, at small mixing angles the 
new resonance 
is predicted \cite{Art2,MPSNODN00} to produce
a much bigger - by a factor of up to $\sim 6$
($\sim 8$),
day-night, or D-N, asymmetry in
the Super-Kamiokande (SNO) sample of solar
neutrino (charged current) events,
whose night fraction is due to the core-crossing 
neutrinos ({\it Core} D-N asymmetry), in comparison with
the asymmetry determined by using the
{\it whole night} 
event sample ({\it Night} asymmetry). As a consequence,
it can be possible to test a substantial part of the
MSW $\nu_e \rightarrow \nu_{\mu (\tau)}$ SMA
solution region in the $\Delta m^2 - \sin^22\theta$
plane (see, e.g., \cite{SKDN98}) by performing 
{\it core} D-N asymmetry measurements 
\cite{Art2,Art1,MPSNODN00,MPDNCoreSK00}.
The Super-Kamiokande collaboration
has already successfully
applied this approach to 
the analysis of their solar neutrino data 
\cite{SKDN98}:
the limit the collaboration has obtained on the
D-N asymmetry utilizing only the 
measured event rate in the night N5 bin, 
approximately 80\% of which is due to the 
Earth-core-crossing solar neutrinos, permitted to
exclude a small part of the MSW SMA solution region.
In contrast, the published Super-Kamiokande upper limit on
the {\it Night} D-N asymmetry \cite{SKDN98}
does not allow to probe the SMA solution region:
the predicted asymmetry is too small (see, e.g., \cite{Art2}).

  The same new enhancement mechanism
can and should be operative also \cite{SP1}
in the $\nu_{\mu} \rightarrow \nu_{e}$
($\nu_{e} \rightarrow \nu_{\mu (\tau)}$)
small mixing angle transitions
of the atmospheric neutrinos crossing the Earth
core if the atmospheric $\nu_{\mu}$ and
$\bar{\nu}_{\mu}$ indeed take part
in large mixing vacuum
$\nu_{\mu}(\bar{\nu}_{\mu}) 
\leftrightarrow \nu_{\tau}(\bar{\nu}_{\tau})$,
oscillations with
$\Delta m^2 \sim (10^{-3} - 8\times 10^{-3})~{\rm eV^2}$,
as is strongly suggested by the Super-Kamiokande
data \cite{SKAtmo98}, and if all
three flavour neutrinos are mixed in vacuum.
The new resonance can take place 
practically for all neutrino
trajectories through the core 
(e.g., for the trajectories with $h = (0^{\circ} - 30^{\circ})$).
It can produce an excess of e-like events
at $-1 \leq \cos\theta_{z}\leq -0.8$,
$\theta_{z}$ being the Zenith angle, 
in the Super-Kamiokande multi-GeV atmospheric
neutrino data and can be responsible 
for at least part of the
strong Zenith angle dependence of the % present in the
Super-Kamiokande multi-GeV and 
sub-GeV $\mu-$like data \cite{SP1,s5398,SPNewE98}.

 The Earth enhancement of the two-neutrino
transitions of interest
of the solar and atmospheric neutrinos
at relatively small mixing angles
has been discussed rather extensively,
see, e.g., refs. \cite{Art1,PastAtmo,PastSun,3nuKP88}.
Some of the articles contain plots
of the probabilities $P_{e2}$ and/or
$P(\nu_{\mu} \rightarrow \nu_{e})$ or
$P(\nu_e \rightarrow \nu_{\mu(\tau)})$
on which one can clearly
recognize now the dominating maximum due to the 
new enhancement effect. 
% (see, e.g., \cite{PastAtmo,PastSun,}).
However, this maximum was invariably interpreted
to be due to the MSW effect in the Earth core
(see, e.g., \cite{PastAtmo,PastSun})
before the appearance of \cite{SP1}.

  In view of the important role the new mechanism
of enhancement of the neutrino transitions in the Earth 
can play in the interpretation of the 
results of the experiments
with solar and atmospheric neutrinos
and in obtaining information about
possible small mixing angle 
$\nu_{\mu}$ ($\bar{\nu}_{\mu}$) and
$\nu_{e}$ ($\bar{\nu}_{e}$) oscillations, 
it would be desirable 
to have an unambiguous understanding of the 
underlying physics of the mechanism.
In \cite{SP1} 
it was interpreted as an effect of 
constructive interference 
between the various probability amplitudes (notably of the 
neutrino transitions in the Earth mantle and in the
Earth core), entering into the sum representing 
the probability amplitude of the transition 
in the Earth. 

    The exact conditions for the enhancement   
of the probabilities $P_{e2}$,
$P(\nu_{\mu} \rightarrow \nu_{e})$
($P(\nu_e \rightarrow \nu_{\mu(\tau)})$, etc.
by the new mechanism
can be studied analytically \cite{SP1} 
owing to the results of a detailed 
numerical analysis \cite{MP98:2layers} 
(see also \cite{3nuKP88}) which showed that
for the calculation of the probabilities 
of interest the two-layer model of 
the Earth density distribution provides 
a very good (in many cases - excellent) 
approximation  
to the more complicated density 
distributions predicted by the existing 
models of the Earth \cite{Stacey:1977,PREM81}.
In the two-layer model,
the neutrinos
crossing the Earth mantle and the core 
traverse effectively
two layers with constant but different densities,
$\bar{\rho}_{man}$ and $\bar{\rho}_{c}$,
and  chemical composition or electron fraction numbers,
$Y_e^{man}$ and $Y_e^{c}$.
The neutrino transitions of interest
\footnote{The densities $\bar{\rho}_{man,c}$
should be considered as mean
densities along the neutrino trajectories.}
result from two-neutrino oscillations
taking place i) first in the mantle 
over a distance $X'$ 
with a mixing angle $\theta'_{m}$ 
and oscillation length $L_{man}$, 
ii) then in the core over a distance $X''$ with 
mixing angle $\theta''_{m} \neq \theta'_{m}$
and oscillation length $L_{c} \neq L_{man}$, 
and iii) again in the mantle over a distance $X'$
with $\theta'_{m}$ and $L_{man}$. 
Utilizing the above prescription, 
analytic expressions for the 
probabilities of neutrino transitions
in the Earth $P_{e2}$,
$P(\nu_{\mu} \rightarrow \nu_{e}) = 
P(\nu_e \rightarrow \nu_{\mu(\tau)})$, etc.
were derived and were
used to study their extrema \cite{SP1}.
Assuming that the neutrino oscillation
parameters $\Delta m^2 > 0$ and $\cos2\theta > 0$ are fixed
and treating the neutrino paths 
$X'$ and $X''$ as independent variables, 
it was found in \cite{SP1} that in addition to the 
maxima corresponding to the MSW effect in the 
Earth mantle or in the Earth core,
there exists a new maximum  
in $P_{e2}$ and
$P(\nu_{\mu} \rightarrow \nu_{e})$.
The latter takes place when the relative phases
the states of 
the two energy-eigenstate neutrinos acquire
after the neutrinos have crossed the mantle,
$\Delta E' X' = 2\pi X'/L_{man}$, and the core,
$\Delta E'' X'' = 2\pi X''/L_{c}$,
obey the constraints,
\begin{equation} 
\Delta E'X' = \pi (2k + 1),~~ \Delta E''X'' = \pi (2k' + 1),
~~k,k' = 0,1,2,...~,  
\end{equation}
\noindent and if the inequality \cite{SP1,SPnu98}  
\begin{equation} 
\cos (2\theta''_{m} - 4\theta'_{m} + \theta)~
< 0,   
\end{equation}       
\noindent is fulfilled. Condition (2) is valid
for the probability $P_{e2}$. 
When equalities (1) hold, (2) ensures that $P_{e2}$
has a maximum.
At the  maximum $P_{e2}$ takes the form \cite{SP1}
\begin{equation} 
P^{max}_{e2} = \sin^2(2\theta''_{m} - 4\theta'_{m}  + \theta). 
\end{equation}
\noindent The analogs of eqs. (1) - (3)
for the probability $P(\nu_{\mu (e)} \rightarrow \nu_{e (\mu;\tau)})$
can be obtained \cite{SP1} by formally setting
$\theta = 0$ while keeping $\theta_{m}' \neq 0$ and
$\theta_{m}'' \neq 0$ in (1) - (3)
\footnote{The conditions for the corresponding maximum
in $P(\nu_{e} \rightarrow \nu_{s})$
and $P(\bar{\nu}_{\mu} \rightarrow \bar{\nu}_{s})$
coincide in form with those for 
$P(\nu_{\mu (e)} \rightarrow \nu_{e (\mu;\tau)})$.}.
The term ``neutrino oscillation
length resonance'' (NOLR) 
was used in \cite{SP1} to denote 
the resonance-type enhancement of $P_{e2}$,
$P(\nu_{\mu} \rightarrow \nu_{e})$ 
($P(\nu_e \rightarrow \nu_{\mu(\tau)})$), etc. 
associated with the conditions (1) - (2).

    The new mechanism of strong enhancement 
of the probabilities 
$P_{e2}$, $P(\nu_{\mu} \rightarrow \nu_{e})$ 
($P(\nu_e \rightarrow \nu_{\mu(\tau)})$), etc. for
the Earth core crossing neutrinos, was identified in 
\cite{SP1} with the neutrino 
oscillation length resonance. This interpretation 
was based on the observation that 
for the several fixed
test values 
of $\sin^22\theta \ltap 0.02$
and Nadir angle $h$ considered in \cite{SP1},
conditions (1) were approximately satisfied 
at the corresponding non-MSW
absolute maxima of $P_{e2}$ and 
$P(\nu_{\mu} \rightarrow \nu_{e})$
in the variable $E/\Delta m^2$ 
and the numerically calculated  
values of $P_{e2}$ and 
$P(\nu_{\mu} \rightarrow \nu_{e})$ 
at these maxima 
were reproduced by eq. (3) 
with a relatively good 
precision. Doubts about the correctness of such an 
interpretation remained since
i) for the parameters of the Earth \cite{Stacey:1977}
(core radius, mantle width, etc.), 
used in the calculations,
the set of conditions (1) - (2),
as was shown in \cite{SP1}, cannot be 
exactly satisfied at
small mixing angles,
and ii) although the phase 
$\Delta E' X'$ was close to $\pi$ at the
relevant % absolute 
maxima of $P_{e2}$ and 
$P(\nu_{\mu} \rightarrow \nu_{e})$
in the test cases studied, 
the values of the phase
$\Delta E'' X''$ differed quite 
substantially from 
those required by eq. (1).
Moreover, the same (or similar) enhancement
mechanism was found to be operative in the
$\nu_2 \rightarrow \nu_{e}$ transitions 
in the case of $\nu_e - \nu_{s}$ mixing and 
in the $\nu_e \rightarrow \nu_{s}$ 
transitions, at small mixing angles 
\cite{SP1}. However,
as was noticed in \cite{SP1}, 
the conditions (1) - (2)
are not even approximately
fulfilled for the $\nu_2 \rightarrow \nu_{e}
\cong \nu_{s} \rightarrow \nu_{e}$ 
and $\nu_e \rightarrow \nu_{s}$ transitions
at small mixing angles; 
similar conclusions were reached for
the $\bar{\nu}_{\mu} \rightarrow \bar{\nu}_{s}$
transitions \cite{s5398}.

   The indicated results stimulated the
systematic study of the extrema of the
probabilities $P_{e2}$,
$P(\nu_{\mu} \rightarrow \nu_{e})$,
$P(\nu_{e} \rightarrow \nu_{s})$, etc.
of the neutrino transitions in the Earth 
and more generally, in a medium of non-periodic 
constant density layers, which is presented here. 
We consider transitions of neutrinos
in a medium 
consisting of i) two layers of different 
constant densities, and ii) three
layers of constant density 
with the first and the third
layers having identical densities and widths 
which differ from those of the second layer.  
Neutrinos pass through such systems of layers
on the way to the neutrino detectors, e.g., 
when they i) are produced in the 
central region of the Earth and traverse 
both the Earth core and mantle,
ii) travel first in vacuum
and then in the Earth mantle,
iii) cross the Earth mantle, the core and the mantle again. 
We derive and analyze 
the complete set of extrema of the two-neutrino 
transition probabilities
in both cases of a medium with two and 
three constant density layers, assuming that \cite{SP1} 
$\Delta m^2/E$ and $\sin^22\theta$
are fixed and treating the 
widths of the layers, $X'$ and $X''$,
as independent variables. 
For both media considered we find that in addition 
to the local maxima associated with 
the MSW effect and the NOLR,
there exist absolute maxima 
caused by a new effect
of enhancement and
corresponding to a total neutrino conversion. We 
find the complete set of such absolute maxima.
The latter are absolute maxima in
any possible independent variables: 
the neutrino energy, 
the widths of one of the layers, etc.  
The conditions for existence
of the new maxima are derived
and it is shown that 
they are fulfilled, in particular,
for the transitions 
of the Earth-core-crossing 
solar and atmospheric neutrinos.
These conditions differ from the
conditions of enhancement
of the probability of transitions
of neutrinos propagating in a medium with
periodically varying density,
discussed in \cite{Param}.
We show that the strong resonance-like 
enhancement of these transitions
is due to the effect of total 
neutrino conversion. 
At small mixing angles and in the case of the
transitions $\nu_2 \rightarrow \nu_{e}$ and
$\nu_{\mu} \rightarrow \nu_{e}$, for instance, 
the values of the parameters $\sin^22\theta$ and $\Delta m^2/E$
at which the total neutrino conversion takes place  
for neutrinos traversing the Earth core 
along a given trajectory
are rather close to the values of the parameters 
for which the NOLR giving 
$P_{e2} = 1$ or $P(\nu_{\mu} \rightarrow \nu_{e}) = 1$
occurs, and the two enhancement 
mechanisms practically coincide. 
For smaller  $\sin^22\theta$
the NOLR reproduces approximately 
(in some cases - roughly) the enhancement: 
for fixed $h$ and  $\sin^22\theta = (10^{-3} - 10^{-2})$, 
for instance, it reproduces the values of 
$P(\nu_2 \rightarrow \nu_{e})$ and
$P(\nu_{\mu} \rightarrow \nu_{e})$  
at the absolute maxima in the 
$\Delta m^2/E$ variable with an error of 
$\sim$ (15 - 60)\% (see further).  
In the case of the 
$\nu_2 \rightarrow \nu_{e} 
\cong \nu_{s} \rightarrow \nu_{e}$,
$\nu_e \rightarrow \nu_{s}$ and 
the $\bar{\nu}_{\mu} \rightarrow \bar{\nu}_{s}$
transitions at small mixing angles,
the total neutrino conversion
occurs for values of the parameters 
for which the NOLR is not realized.
In all transitions, however,
only the total neutrino 
conversion mechanism is operative.
This mechanism is responsible also for the 
strong enhancement of the 
above transitions at large mixing angles as well as of 
the $\nu_2 \rightarrow \nu_{e}$ transitions due to 
$\nu_{e} - \nu_{s}$ mixing and of the 
$\nu_e \rightarrow \nu_{s}$ and 
the $\bar{\nu}_{\mu} \rightarrow \bar{\nu}_{s}$
transitions at small and large mixing angles. 

  We demonstrate that the 
NOLR and the newly found 
resonance-like enhancement 
effect are caused by a maximal constructive
interference between the amplitudes of the neutrino
transitions in the different density layers.
Thus, the local and absolute maxima they produce in 
the neutrino transition
probabilities are of interference nature. 
Therefore we confirm the physical
interpretation of the strong resonance-like enhancement
of the transitions in the Earth of the Earth-core-crossing
solar and atmospheric neutrinos, given in \cite{SP1}.

  Most of the results of our analysis
are illustrated by the physically
realistic
examples of the transitions of 
neutrinos passing through the Earth. 
These examples and the results regarding the 
transitions of Earth-core-crossing
solar and atmospheric
neutrinos are derived using the Stacey
model from 1977 \cite{Stacey:1977} as a
reference Earth model. 
The density distribution
in the Stacey model is
spherically symmetric
and in addition to the two major density structures -
the core and the mantle, there are 
a certain number of substructures (shells or layers).
The core has a radius $R_c = 3485.7~$km,
the Earth mantle depth is approximately $R_{man} = 2885.3~$km,
and the Earth radius in the Stacey model
is $R_{\oplus} = 6371~$km.
Therefore, when
the Nadir angle $h$ is less than 33.17$^\circ$, 
neutrinos arriving at the
detector pass through three layers: 
the mantle, the core and the mantle again.
The mean values of the matter densities of the core and of the mantle
read, respectively: $\bar{\rho}_c \cong 11.5~{\rm g/cm^3}$ and
$\bar{\rho}_{man} \cong 4.5~{\rm g/cm^3}$.
The density distribution in the 1977 Stacey model
practically coincides with that 
in the more recent PREM model \cite{PREM81}.
For $Y_e$ we have used the standard
values \cite{Stacey:1977,PREM81,CORE}
(see also \cite{Art2})
$Y_e^{man} = 0.49$ and $Y_e^{c} = 0.467$.

  A brief description of the results of the present study is given in
ref. \cite{ChPet99S}.

\section{Preliminary Remarks} 

\indent  We will consider the simple case of 
mixing of two weak-eigenstate neutrinos $\nu_{\alpha}$ 
and $\nu_{\beta}$, $\alpha \neq \beta = e,\mu,\tau,s$,
$\nu_s$ being a sterile neutrino, in vacuum.
The mixing matrix relating the states of the neutrinos 
$\nu_{\alpha,\beta}$ % and $\nu_{\beta}$ 
and of the neutrinos $\nu_{1,2}$ % and $\nu_2$ 
having definite masses $m_{1,2}$ % and $m_2$ 
in vacuum is chosen in the form:
\begin{equation}
\left(\begin{array}{c} \nu_{\alpha} \\ \nu_{\beta} \end{array}\right) =
\left(\begin{array}{cc} \cos\theta & \sin\theta \\
-\sin\theta & \cos\theta \end{array}\right)
\left(\begin{array}{c} \nu_1 \\ \nu_2 \end{array}\right),
\end{equation}

\noindent where 
$\theta$ is the vacuum mixing angle. In the
case of relativistic and stable neutrinos $\nu_{1,2}$, 
the evolution equation for neutrino
propagation in constant-density medium reads
\begin{equation}
i\frac{{\rm d}}{{\rm d}t}
\left(\begin{array}{c} \nu_{\alpha} \\ \nu_{\beta}\end{array}\right) =
{\Delta E\over 2} 
\left(\begin{array}{cc} -\cos(2\theta_m) & \sin(2\theta_m) \\
\sin(2\theta_m) & \cos(2\theta_m) \end{array}\right)
\left(\begin{array}{c} \nu_{\alpha} \\ \nu_{\beta}\end{array}\right)=
{\cal M} \left(\begin{array}{c} \nu_{\alpha} \\ \nu_{\beta}\end{array}\right).
\end{equation}

\noindent Here $\theta_m$ is the mixing angle in matter, 
\begin{equation}
\cos(2\theta_m) = {1\over \Delta E}~((\Delta m^2/2E) \cos(2\theta) - 
          V_{\alpha\beta}), 
\end{equation}

\noindent $E$ being the neutrino energy, 
$\Delta E$ is the difference between the energies of the two 
energy-eigenstate neutrinos in matter,
\begin{equation}
\Delta E = {\Delta m^2 \over 2E} \sqrt{\left(\cos(2\theta)-
{2EV_{\alpha\beta} \over \Delta m^2} \right)^2 + \sin^2(2\theta)},
\end{equation} 

\noindent $\Delta m^2 = m_2^2-m_1^2$, 
and $V_{\alpha \beta}$ is the difference between the
effective potentials of $\nu_{\alpha}$ and $\nu_{\beta}$  
in the medium. We shall always assume in what follows that
\begin{equation}
\cos(2\theta) > 0,~~~ \Delta m^2 > 0.
\label{cosdelta}
\end{equation} 

\noindent In the case of neutrino propagation in an
electrically neutral unpolarized cold medium, like the Earth,
one has:
\begin{equation}
V_{e\mu} = \sqrt{2}G_F N_e,~~V_{es} = \sqrt{2}G_F (N_e - {1\over 2}N_n),~~ 
V_{\mu s}=-\sqrt{2}G_F N_n/2,
\end{equation}

\noindent where $N_e$ and $N_n$ are the electron and the neutron 
number densities in the medium, respectively. The antineutrino states
$\bar{\nu}_{\alpha}$ and  $\bar{\nu}_{\beta}$ satisfy 
the same equations (5)
with $V_{\alpha\beta}$ replaced by $V_{\bar{\alpha} \bar{\beta}} = 
- V_{\alpha\beta}$ in eqs. (6) and (7). For the Earth $N_e \cong N_n$ and
in addition to $V_{e\mu} > 0$ we have $V_{es} > 0$ and   
$V_{\bar{\mu} \bar{s}} > 0$. If (8) holds, the 
neutrino mixing can be enhanced by the Earth matter only in the cases of 
$\nu_{\mu}~(\nu_e) \rightarrow \nu_e~(\nu_{\mu;\tau})$
($\nu_2 \rightarrow \nu_e$),
$\nu_e \rightarrow \nu_s$ and
$\bar{\nu}_{\mu} \rightarrow \bar{\nu}_s$ transitions, while
for, e.g., $\cos(2\theta) > 0$ and $\Delta m^2 < 0$
this will be true for the 
$\bar{\nu}_{\mu}~(\bar{\nu}_e) \rightarrow \bar{\nu}_e~(\bar{\nu}_{\mu;\tau})$
($\bar{\nu}_2 \rightarrow \bar{\nu}_e$),
$\bar{\nu}_e \rightarrow \bar{\nu}_s$ and
$\nu_{\mu} \rightarrow \nu_s$ transitions. 
Our general results will be 
formulated for the generic weak-eigenstate neutrino
transitions, $\nu_{\alpha} \rightarrow \nu_{\beta}$;
for neutrinos crossing the Earth we will be 
interested either in 
the former or in the latter set of transitions,
depending on whether eq. (8) holds or 
$\Delta m^2 \cos(2\theta) < 0$. The case of
(solar) $\nu_{2} \rightarrow \nu_e$  
transitions in a three-layer medium 
(the Earth) will be considered separately. 

  The evolution of the neutrino system 
is given by the unitary matrix (see, e.g., \cite{JJSaku85,PCWKim93})
\begin{equation}
U=\exp(-i{\cal M}t)=\cos\phi-i(\mbox{\boldmath ${\sigma n}$})\sin\phi,
\label{evolution}
\end{equation}

\noindent where 
\begin{equation}
\phi=\frac{1}{2}\Delta E~t
\end{equation}
\noindent and 
\begin{equation}
{\bf n}=(\sin(2\theta_m),0,-\cos(2\theta_m))
\end{equation}

\noindent is a real unit vector.
The probabilities of the $\nu_{\alpha} \to \nu_{\beta}$ 
($\nu_{\beta} \to \nu_{\alpha}$) transition, 
$P_{\alpha \beta (\beta \alpha)} = |A_{\alpha \beta (\beta \alpha)}|^2$,
and of the $\nu_{\alpha}$ ($\nu_{\beta}$) survival, 
$P_{\alpha \alpha (\beta \beta)} = |A_{\alpha \alpha (\beta \beta)}|^2$,  
are determined by the elements of the evolution
matrix $U$, which coincide with the four different probability
amplitudes $A_{\alpha \beta} = U_{\beta \alpha}$, etc.:
\begin{equation}
P_{\alpha \beta} = |U_{\beta \alpha}|^2=
\left(n_1\sin\phi\right)^2 + \left(n_2\sin\phi\right)^2 = P_{\beta \alpha}
\label{Pem}
\end{equation}

\noindent and
\begin{equation}
P_{\alpha \alpha} = |U_{\alpha \alpha}|^2=
\cos^2\phi + \left(n_3\sin\phi\right)^2 = 
1 - P_{\alpha \beta} = P_{\beta \beta}.
\label{Pee}
\end{equation}

  We are interested in the extrema of the 
transition probability
$P_{\alpha \beta}$ when neutrinos
propagate through a medium with 
(nonperiodic sequence of) layers of
different constant density and chemical composition. In this case 
the evolution matrix, as is well-known, 
represents a product of the evolution matrices for
the different layers and can always be written in the same form as in eq. 
(\ref{evolution}). It follows from (\ref{Pem}) and (\ref{Pee}) that the
conditions for 
an {\it absolute} maximum of the transition probability,
$P_{\alpha \beta} = 1$, read
\begin{equation}   
max~P_{\alpha \beta} = 1:~\left\{ \begin{array}{l} \cos\phi=0 \\
n_3\sin\phi=0, \end{array} \right.
\label{max}
\end{equation}
\noindent while the conditions for % the presence of 
an {\it absolute} minimum of $P_{\alpha \beta}$ % , $P_{\alpha \beta} = 0$, 
have the form:
\begin{equation}
min~P_{\alpha \beta} = 0:~\left\{ \begin{array}{l} n_1\sin\phi=0 \\
n_2\sin\phi=0. \end{array} \right.
\label{min}   
\end{equation}

   When neutrinos propagate in a constant-density medium the
transition probability 
\begin{equation}
P_{\alpha \beta} = {1\over2}(1 - \cos 2\phi)\sin^2(2\theta_m) = 
 (\sin \phi \sin (2\theta_m))^2
\label{MSW}
\end{equation}

\noindent can be non-negligible 
even in the case of small 
vacuum mixing angle $\theta$
if the neutrino mixing in matter is maximal, $\theta_m=\pi/4$. 
This is the well-known MSW effect 
which takes place when the resonance
condition 
\begin{equation}
\cos(2\theta_m) = 0
\label{res}
\end{equation}

\noindent is fulfilled. 
In order to get 
total neutrino conversion,
$P_{\alpha \beta} = 1$,
the additional requirement 
\begin{equation}
\cos\phi = 0
\label{fmax}
\end{equation}

\noindent has to be satisfied. 
For a fixed time of propagation $t$, or a distance
$X \cong t$ traveled by the neutrino, 
there always exists a solution of the system
of equations (\ref{max}) 
The {\it absolute} minima
$P_{\alpha \beta} = 0$ correspond, as it follows from (\ref{min}), 
to the curves 
\begin{equation}
\sin\phi=0.
\label{fmin}
\end{equation}

  In the next sections we discuss the 
case of neutrino propagation through
a number of nonperiodic constant-density layers. 
A new phenomenon of maximal constructive 
interference between the probability  
amplitudes of the transitions in the 
different layers takes place in this case,
leading to a substantial enhancement of 
the transition probability. 
Even when the oscillation 
parameters have values very different from those required by 
the resonance conditions (\ref{res})
for the individual layers, the neutrino 
transition probability 
can reach its absolute
maximum, $P_{\alpha \beta} = 1$, due to this effect.

\section{Medium with Two Constant - Density Layers}

\indent Suppose neutrinos $\nu_{\alpha}$ or $\nu_{\beta}$ have crossed 
a medium consisting of two layers 
having constant but different density and chemical composition.
These could be the Earth core and mantle (for neutrinos born in 
the Earth central region),
the vacuum and the Earth mantle, etc.
Let us denote the effective potential differences and 
the lengths of the paths of the neutrinos (antineutrinos) 
in the two layers by  
$V'_{\alpha \beta}$ ($V'_{\bar{\alpha} \bar{\beta}}$), 
$V''_{\alpha \beta}$ ($V''_{\bar{\alpha} \bar{\beta}}$)
and $X'$, $X''$,
respectively. We shall assume without loss of generality that
\begin{equation} 
 0 \leq  |V'_{\alpha \beta (\bar{\alpha} \bar{\beta})}| 
< |V''_{\alpha \beta (\bar{\alpha} \bar{\beta})}|.
\label{poten}
\end{equation}

\noindent The transitions
of $\nu_{\alpha}$ and $\nu_{\beta}$ 
crossing the two layers will be
determined by the two sets of two parameters - 
the mixing angle in matter (6) or the unit vector (12), and the 
phase difference (11) with $t = X'$ or $X''$,
which characterize the 
transitions in each of the layers: 
$\theta'_m$ or ${\bf n}'$ and $\phi'$,  and 
$\theta''_m$ or ${\bf n}''$ and $\phi''$.  
Since the evolution matrix of the system is given by $U = U''U'$, $U'$ and
$U''$ being the evolution matrices in the first and in the second layers,
the parameters of $U$,
$\Phi$ and ${\bf n}$ (eq. (\ref{evolution})),
can be expressed in terms of 
$\phi'$, ${\bf n}'$  and $\phi''$, ${\bf n}''$ as follows 
(see, e.g., \cite{JJSaku85}):
\begin{equation}
\left\{ 
\begin{array}{l}
\cos\Phi=\cos\phi'\cos\phi''-({\bf n}'\cdot{\bf n}'')\sin\phi'\sin\phi''\\
= \cos\phi'\cos\phi''-\cos(2\theta_m'' - 2\theta_m')\sin\phi'\sin\phi'',\\
{\bf n}\sin\Phi={\bf n}'\sin\phi'\cos\phi''+{\bf n}''\cos\phi'\sin\phi''
-[{\bf n}'\times{\bf n}'']\sin\phi'\sin\phi''.
\end{array} \right.
\label{par-2}
\end{equation}

\noindent  Using, e.g., eqs. (\ref{Pem}) and (\ref{par-2}) 
it is not difficult to derive
the transition probability $P_{\alpha \beta} = P_{\beta \alpha}$ 
in this case:
$$ P_{\alpha \beta} 
 = {1\over {2}}\left [1 - \cos 2\phi' \right ] \sin^2 2\theta'_{m}
+ {1\over {2}}\left [1 - \cos 2\phi'' \right ] \sin^2 2\theta''_{m}
+ {1\over {4}} \left [1 - \cos 2\phi'\right ]
\left [1 - \cos 2\phi''\right ] \times$$
\begin{equation}
\times \left [ \sin^2(2\theta''_{m} 
    - 2\theta'_{m}) - \sin^2 2\theta'_{m} -
 \sin^2 2\theta''_{m} \right ]
+ ~{1\over {2}} \sin 2\phi'~\sin 2\phi''~\sin 2\theta'_{m} 
\sin 2\theta''_{m}. 
\label{Pem-2}
\end{equation}
\noindent Due to T-invariance and unitarity, the probability (\ref{Pem-2}) is
symmetric with respect to the interchange of the parameters of the two layers,
i.e., does not depend on the order in which the neutrinos traverse them.

  For fixed $\sin^2(2\theta)$ and $\Delta m^2/E$ 
and given density and chemical composition in the two layers, 
the unit vectors ${\bf n}'$ and ${\bf n}''$ are also fixed for each layer.
The phases $\phi'$ and $\phi''$,
which depend on the widths of the layers $X'$ and $X''$,
can be independent variables of the
system  if the two widths $X'$ and $X''$ 
can be varied independently. 
This is an interesting case with a clear physical meaning 
and we are going to investigate it in the present and the 
next Sections.  In this way the NOLR resonance 
in the $\nu_2 \rightarrow \nu_e$,
$\nu_{\mu}~(\nu_e) \rightarrow \nu_e~(\nu_{\mu;\tau})$
and $\nu_e \rightarrow \nu_s$ transitions
was discovered~\cite{SP1}.
Let us note that in the case of the
Earth, $X'$ and $X''$ cannot be treated as independent
variables as long as the Earth radius and the Earth core 
radius are fixed. Neutrino transitions in the Earth 
will be considered at the end of this Section 
and in Section 6.

   Varying the phases $\phi'$ and $\phi''$
we can investigate the number and the 
structure of the extrema of the 
transition probability $P_{\alpha \beta}$. 
These are determined by the system of two equations
\begin{equation}
{dP_{\alpha \beta}\over {d\phi'}} = \sin 2\theta'_{m}~
F(2\theta'_{m} - 2\theta''_{m},2\theta''_{m}; 2\phi',2\phi'') = 0,
\label{cond1}
\end{equation} 
\begin{equation}
{dP_{\alpha \beta}\over {d\phi''}} = \sin 2\theta''_{m}~
F(2\theta''_{m} - 2\theta'_{m},2\theta'_{m}; 2\phi'',2\phi') = 0,
\label{cond2}
\end{equation} 
\noindent where
\begin{equation}
F(Y,Z;\varphi,\psi) = \sin Y \cos Z \sin \varphi + \sin Z \left [ \sin \varphi
\cos \psi \cos Y + \cos \varphi \sin \psi \right ],
\label{F}
\end{equation} 

\noindent and by two known supplementary conditions 
on the values of the second
derivatives of $P_{\alpha \beta}$  at the points where eqs. (\ref{cond1})
and (\ref{cond2}) hold.

  We are interested first of all in the {\it absolute} maxima of the 
neutrino conversion,
\begin{equation} 
case~A:~~~max~P_{\alpha \beta} = 1.
\end{equation}

\noindent They are determined 
by the equations (\ref{max}) and (\ref{par-2}) 
\begin{equation}
max~P_{\alpha \beta} = 1:~\left\{ \begin{array}{l}
\cos\Phi \equiv \cos\phi'\cos\phi''-\cos(2\theta_m'' - 
2\theta_m')\sin\phi'\sin\phi'' = 0 \\
n_3\sin\Phi \equiv - [
\cos(2\theta_m')\sin\phi'\cos\phi''+\cos(2\theta_m'')\cos\phi'\sin\phi''] = 0.
\end{array} \right.
\label{maxAc}
\end{equation}

\noindent It is not difficult to check that conditions (\ref{cond1}) and 
(\ref{cond2}) are satisfied if relations (\ref{maxAc}) hold.
The solutions of (\ref{maxAc}) can be readily found: 
\begin{equation}
solution~A:~\left\{ \begin{array}{l}
\tan\phi'=\pm\sqrt{{\displaystyle -\cos(2\theta_m'')\over
\displaystyle\cos(2\theta_m')\cos(2\theta_m'' - 2\theta_m')}}~, \\
\tan\phi''=\pm\sqrt{{\displaystyle -\cos(2\theta_m')\over
\displaystyle\cos(2\theta_m'')\cos(2\theta_m'' - 2\theta_m')}}~,
\end{array} \right.
\label{max-2As}
\end{equation}

\noindent where the signs are correlated. Obviously, the 
solutions (\ref{max-2As}) do not exist for 
any pairs of values of the parameters 
$\sin^2 (2\theta)$ and $\Delta m^2/E$. 
%%%%%%%%%%%%%%%%%%%%%%%%%%%%%%%%%%%
\begin{figure}[t]
\begin{center}
\includegraphics[width=14cm,height=10cm]{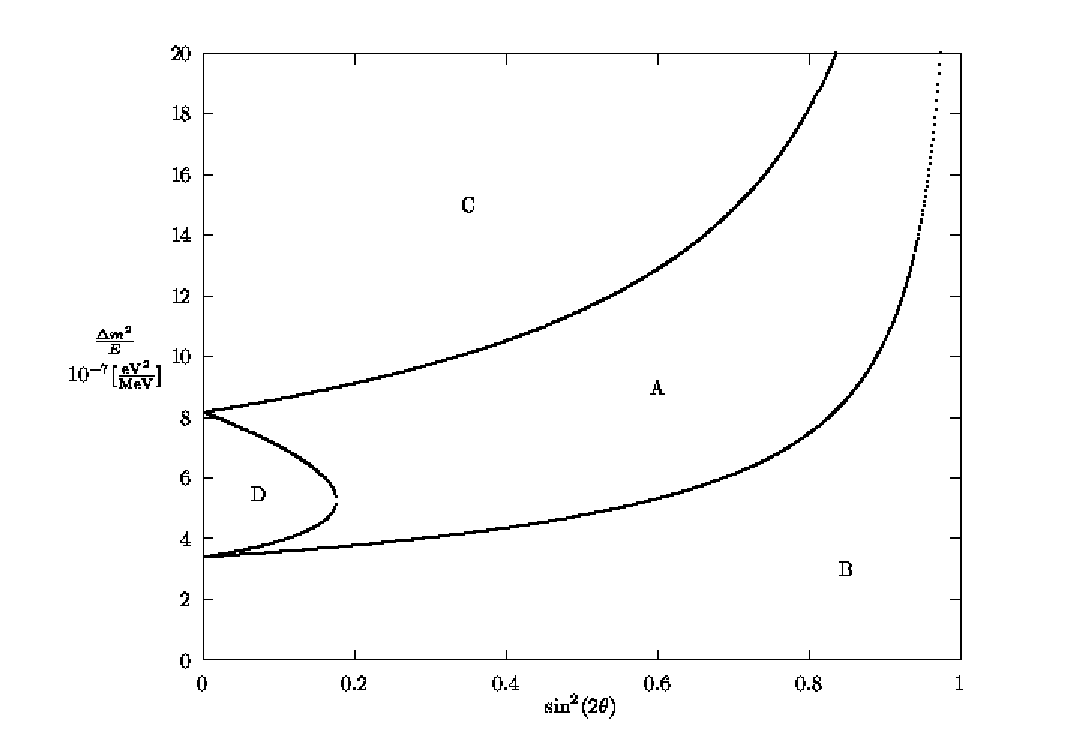}
\caption{
The regions of
the four different solutions $A$, $B$, $C$ and $D$
(eqs. (\ref{max-2As}), (\ref{B}), (\ref{C}) and 
(\ref{D})) for the maxima of the
transition probability $P_{e\mu} = P_{\mu e}$
in a two-layer medium. The two different layers
correspond to the core and the mantle of the Earth
\protect\cite{Stacey:1977,PREM81}.
}
\label{Fig1}
\end{center}
\end{figure}
%%%%%%%%%%%%%%%%%%%%%%%%%%%%%%%%%%%%%%%%%
%
\noindent Under the assumptions (\ref{cosdelta}), (\ref{poten})
and for $V'_{\alpha \beta} \geq 0$ (e.g., for the 
$\nu_{\mu}~(\nu_e) \rightarrow \nu_e~(\nu_{\mu;\tau})$,
$\nu_e \rightarrow \nu_s$ and
$\bar{\nu}_{\mu} \rightarrow \bar{\nu}_s$ 
transitions of the Earth-crossing neutrinos), 
they can take place only in the
region limited by the three  
conditions
\footnote{These conditions can also be derived from the 
two supplementary inequalities ensuring that the 
solutions (\ref{max-2As}) of (\ref{cond1}) and (\ref{cond2}) 
indeed correspond to maxima of $P_{\alpha \beta}$.} 
(Fig. 1)
\begin{equation}
region~A:~\left\{ \begin{array}{l}
\cos 2\theta_m' \ge 0 \\
\cos 2\theta_m'' \le 0 \\
\cos(2\theta_m''-2\theta_m') \ge 0.
\end{array} \right.
\label{regionA}
\end{equation}

   The first boundary, $\cos 2\theta_m' = 0$, corresponds to a 
{\it total} neutrino conversion  {\it in the first layer}. 
From (\ref{max-2As}) we get the correct
result for the relevant conditions on the phases
(see (\ref{fmax}) and (\ref{fmin})),
\begin{equation}
solution~B:~\left\{ \begin{array}{l}
\cos\phi'=0,~or~ 2\phi' = \pi (2k' + 1),~k' = 0,1,..., \\
\sin\phi''=0,~or~ 2\phi'' = 2\pi k'',~k'' = 0,1,...,
\end{array} \right.
\label{B}
\end{equation}

\noindent guarantying that i) neutrino conversion 
takes place in the first layer, and ii) no neutrino conversion 
occurs in the second layer, so that when 
$\cos 2\theta_m' = 0$ we have
maximal transition probability, $P_{\alpha \beta} = 1$. 
It is easy to show by direct calculations using eqs.
(\ref{Pem-2}) - (\ref{cond2}) that the phase 
conditions (\ref{B}) correspond to the
local maxima
\begin{equation}
case~B:~ max~P_{\alpha \beta} = \sin^2 2\theta_m',
\end{equation}

\noindent provided the oscillation parameters 
$\sin^2(2\theta)$ and $\Delta m^2/E$
belong to the region $B$,
\begin{equation}
region~B:~\cos 2\theta_m' \leq 0.
\label{regionB}
\end{equation}

 It proves useful to interpret this result in terms of
probability amplitudes. If neutrinos $\nu_{\alpha}$ have crossed
the two layers, the amplitude of the 
$\nu_{\alpha} \rightarrow \nu_{\beta}$ transition 
represents a sum of two terms: 

\begin{equation}
A_{\alpha \beta} = U_{\beta \alpha} = 
U''_{\beta \alpha}U'_{\alpha \alpha} + 
U''_{\beta \beta}U'_{\beta \alpha},
\label{amplitude}
\end{equation}

\noindent where
\begin{equation}
\begin{array}{ll}
U'_{\alpha \alpha } = \cos\phi'+i\cos(2\theta_m')\sin\phi', &
U''_{\beta \alpha} = -i\sin(2\theta_m'')\sin\phi'', \\
U'_{\beta \alpha} = -i\sin(2\theta_m')\sin\phi', &
U''_{\beta \beta} = \cos\phi''-i\cos(2\theta_m'')\sin\phi''.
\end{array}
\label{amplitudes}
\end{equation}

\noindent  In the general case the transition probability 
\begin{equation}
P_{\alpha \beta} = |U''_{\beta \alpha}|^2 |U'_{\alpha \alpha}|^2 + 
|U''_{\beta\beta}|^2 |U'_{\beta \alpha}|^2
+ 2{\rm Re}\left(\left(U''_{\beta \alpha}U'_{\alpha \alpha}\right)^* 
U''_{\beta \beta}U'_{\beta \alpha}\right)
\label{Pampl}
\end{equation}  

\noindent is a sum of two products of the probabilities of
neutrino oscillations in the different layers and of 
the interference term. The latter can
play a crucial role in the resonance-like 
enhancement of the neutrino transitions. 

  The phase requirements (\ref{B}) reduce the expression in   
eq. (\ref{Pampl}) to only one 
term, $P_{\alpha \beta} = |U'_{\beta \alpha}|^2$, with no 
contribution from the interference term because $U''_{\beta \alpha} = 0$.
Thus, the maxima in region $B$ (\ref{regionB}) can be ascribed to
neutrino transitions taking place in the first layer only.
The absolute maxima, 
max $P_{\alpha \beta} = 1$, correspond for 
$\sin^22\theta < 1$ to
the MSW effect in the first layer. 
 
  The second boundary $\cos(2\theta_m'') = 0$ in (\ref{regionA})
corresponds to
a {\it total} neutrino conversion in the second layer,
\begin{equation}
solution~C:~\left\{ \begin{array}{l}
\sin\phi' = 0,~or~2\phi' = 2\pi k',~k' = 0,1,..., \\
\cos\phi'' = 0,~or~ 2\phi'' = \pi (2k'' + 1),~k'' = 0,1,...~.
\end{array} \right.
\label{C}
\end{equation}
 
\noindent This case is completely
analogous to the previously considered 
case $B$, with the two layers interchanged.
Using eqs. (\ref{Pem-2}) - (\ref{cond2}) 
we get solution $C$
\noindent which realizes the {\it local} maxima
\begin{equation}
case~C:~~ max~P_{\alpha \beta} = \sin^2 2\theta_m'',
\end{equation}

\noindent in the region
\begin{equation}
region~C:~\cos 2\theta_m'' \ge 0.
\label{regionC}
\end{equation} 

\noindent The case of equality in eq. (\ref{regionC}) % This 
can be associated with
the MSW effect taking place in the second layer. 

  We get a very different mechanism of 
enhancement of the neutrino transition 
probability $P_{\alpha \beta}$
in the intermediate region
\begin{equation}
\left\{ \begin{array}{l}
\cos 2\theta_m' > 0, \\
\cos 2\theta_m'' < 0, \\
\end{array} \right.
\label{intermediate}
\end{equation}

\noindent located between 
the regions $B$ (\ref{regionB}) and
$C$ (\ref{regionC}).
The maxima of $P_{\alpha \beta}$ in this region 
are caused by a maximal contribution 
of the interference term 
in eq. (\ref{Pampl}) for $P_{\alpha \beta}$.
Indeed, using eq. (\ref{amplitude}) one can write
the probability $P_{\alpha \beta}$ in the form:
\begin{equation}
P_{\alpha \beta} = |{\bf z}_1 + {\bf z}_2|^2, 
\label{Pzet}
\end{equation}

\noindent where ${\bf z}_1 =
({\rm Re}(U''_{\beta \alpha}U'_{\alpha \alpha}),
{\rm Im}(U''_{\beta \alpha}U'_{\alpha \alpha}))$ and 
${\bf z}_2 = ({\rm Re}(U''_{\beta \beta}U'_{\beta \alpha}),
{\rm Im}(U''_{\beta \beta}U'_{\beta \alpha}))$ 
are two vectors in the complex plane.
In the case of the probability $P_{\alpha \beta}$ 
the interference is maximal and
constructive, as it follows from (\ref{Pzet}),
when ${\bf z}_1$ and ${\bf z}_2$ are collinear and
point in the same direction, i.e., when 
\begin{equation}
{{\rm Im}(U''_{\beta \alpha}U'_{\alpha \alpha}) 
     \over {\rm Im}(U''_{\beta \beta}U'_{\beta \alpha})} =
{{\rm Re}(U''_{\beta \alpha}U'_{\alpha \alpha}) 
\over {\rm Re}(U''_{\beta \beta}U'_{\beta \alpha})} > 0.
\label{maxint}
\end{equation}

\noindent 
The second equation in (\ref{max}) (or (\ref{maxAc})) ensures that 
the two vectors are collinear, while the constraints 
(\ref{intermediate}) 
guarantee that they point in the same direction.
Actually, the equation in (\ref{maxint}) coincides 
with the second equation in
(\ref{max}) for the resonance condition (\ref{res})
\begin{equation}
\cos(2\theta_m) = {1\over \sin\phi }
\left [ \cos(2\theta')\sin\phi'\cos\phi'' + 
\cos(2\theta'')\cos\phi'\sin\phi'' \right ] = 0,
\label{res-2}
\end{equation}

\noindent with the additional constraints (\ref{intermediate}), 
corresponding to the intermediate region where the new
type of enhancement takes place. 
The transition probability of interest 
(\ref{Pem-2}) can obviously 
be represented in the two 
layer case under discussion 
also in the form (\ref{MSW}),
\begin{equation}
P_{\alpha \beta} = \sin^2(2\theta_m)\sin^2\phi,
\end{equation}

\noindent where the first and the second multipliers are defined by
eq. (\ref{res-2}) and the first equation in (\ref{par-2}).
To get a total neutrino conversion, $P_{\alpha \beta} = 1$, 
not only eq. (\ref{res-2}) has to hold, but
also  condition (\ref{fmax}) must be satisfied. 
This is possible in region $A$ (\ref{regionA}), 
but not in the whole intermediate region
(\ref{intermediate}).

 Indeed, the boundary $\cos(2\theta_m''- 2\theta_m') = 0$ 
in (\ref{regionA}) corresponds to a 
maximum of $P_{\alpha \beta}$ associated
with the phase constraints
\begin{equation}
solution~D:~\left\{ \begin{array}{l}
\cos\phi'=0,~or~ 2\phi' = \pi (2k' + 1),~k' = 0,1,..., \\
\cos\phi''=0,~or~ 2\phi'' = \pi (2k'' + 1),~k'' = 0,1,...~ .
\end{array} \right.
\label{D}
\end{equation}

\noindent  The above conditions are necessary
for a maximal neutrino conversion in each layer 
(see (\ref{fmax}) and  (\ref{amplitudes})). As can be shown 
exploiting eqs. (\ref{Pem-2}) - (\ref{cond2}), they lead to 
local maxima
\begin{equation}
case~D:~~ max~P_{\alpha \beta} = \sin^2(2\theta_m''-2\theta_m'),
\end{equation}

\noindent if the oscillation parameters belong to 
the finite region $D$,
\begin{equation}
region~D:~\cos(2\theta_m''-2\theta_m')\leq 0.
\label{regionD} 
\end{equation}

   Our results show a very interesting 
feature of the neutrino transitions 
in a medium consisting of 
two constant-density layers. One can have  
total neutrino conversion,
$P_{\alpha \beta} = 1$, even if the 
MSW resonance does not take place
in any of the two layers. As we have seen, this is possible in
region $A$, excluding the two boundaries $\cos(2\theta'_m) = 0$
and $\cos(2\theta''_m) = 0$, corresponding to the MSW effect.
Therefore, in order to have 
a large transition probability, $P_{\alpha \beta} \cong 1$,
a periodic density profile is not required even when 
the MSW resonance is not realized in any of the two layers. 
These conclusions remain valid, as we shall see,
for transitions of neutrinos traversing 
three layers of constant density, 
e.g., for the transitions of
the solar and atmospheric neutrinos 
crossing the Earth core on the way to the detectors.
It should be emphasized that solution 
$A$, eq. (\ref{max-2As}), providing the 
absolute maxima of $P_{\alpha \beta}$,  
can be and was derived directly from 
eqs. (15) and (22) 
without utilizing the standard extrema
equations (24) - (25), i.e., {\it its derivation 
does not depend on the chosen set of  
variables} of $P_{\alpha \beta}$, 
which are varied 
to obtain the extrema conditions. 
Obviously, it can also be obtained
for {\it a given set of variables}
from eq. (23) and the corresponding 
extrema conditions of the type of  (24) - (25). 

  The {\it absolute} minima of $P_{\alpha \beta}$
in the case under discussion are
located on the intersections of the curves
\begin{equation}
min~P_{\alpha \beta} = 0:~\left\{ \begin{array}{l}
\sin\phi'=0,~or~2\phi' = 2\pi k',~k' = 0,1,..., \\
\sin\phi''=0,~or~ 2\phi'' = 2\pi k'',~k'' = 0,1,...~.
\end{array} \right.
\end{equation}

  Our results apply to the case 
when the first layer is vacuum $V' = 0$
(Fig. 2), because there is
no principal difference between the neutrino oscillations 
in constant-density medium and in vacuum.
If we assume the nonzero density layer to be the Earth mantle,
maximal neutrino conversion (solution A) can take place provided
neutrinos travel a distance of $\sim$ few$\times 10^{3}$ km in vacuum. 
%%%%%%%%%%%%%%%%%%%%%%%%%%%%%%%%%%%
\begin{figure}[t]
\begin{center}
\includegraphics[width=14cm,height=10cm]{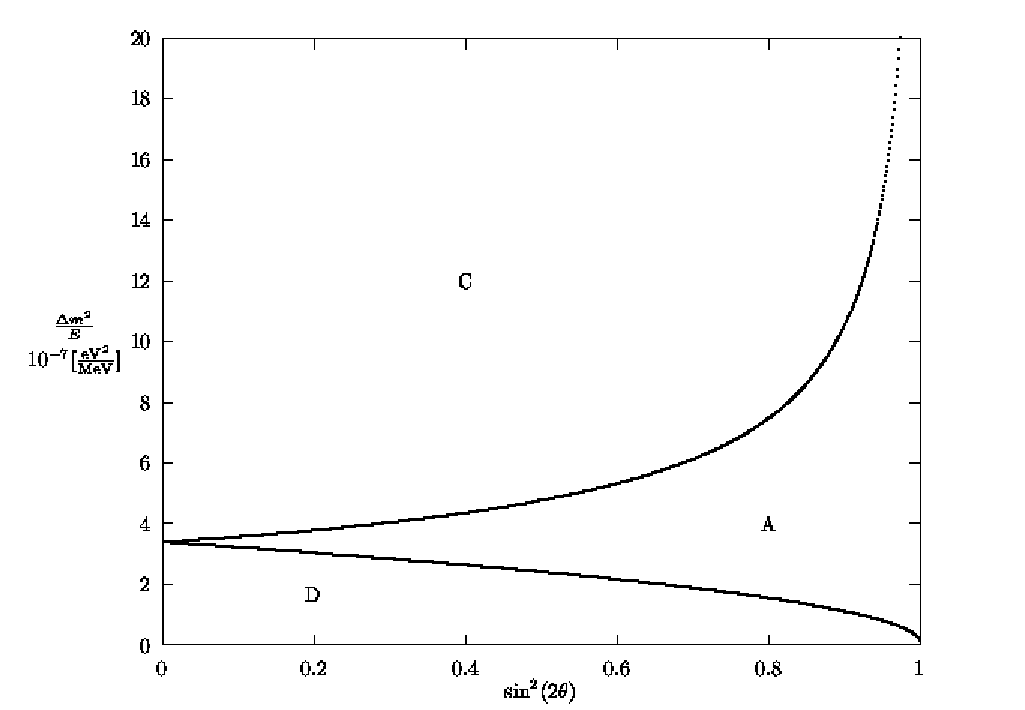}
\caption{The same as in figure 1
for the case vacuum - Earth mantle.
}
\label{Fig2}
\end{center}
\end{figure}
%%%%%%%%%%%%%%%%%%%%%%%%%%%%%%%%%%%%%%%%%
%

 The above results are valid also, e.g., 
for neutrinos born in the central 
region of the Earth. Such neutrinos cross two layers -
the Earth core and mantle, on the way to the Earth surface.
If the dimensions of the region of neutrino production are
small compared to the Earth core radius, the neutrino path 
lengths $X'$ and $X''$ are fixed. 
The extrema condition with respect to the variable $E$
is given in ref. \cite{ChPet99S}. 
It differs from the the extrema conditions
(\ref{cond1}) and 
(\ref{cond2}). Accordingly, the solutions of the system 
(\ref{cond1}) - (\ref{cond2})
do not correspond, in general, to extrema in the variable $E$. 
More specifically, solutions $B$, $C$ and $D$, corresponding 
to local maxima of $P_{\alpha \beta}$,
$max~P_{\alpha \beta} < 1$, in the variables
$\phi'$  and $\phi''$, do not provide
local maxima in the variable $E$.  
Only solution $A$ provides the absolute maxima 
of the neutrino transition probabilities of interest, 
$P_{\alpha\beta} = 1$, in any variable.
Solution $A$ is realized 
for $\nu_{\mu}~(\nu_e) \rightarrow \nu_e~(\nu_{\mu;\tau})$,
$\nu_e \rightarrow \nu_s$ and 
$\bar{\nu}_{\mu} \rightarrow \bar{\nu}_{s}$
transitions of neutrinos born in the 
central region of the Earth, as Table 1 demonstrates.
% This solution has also a rather simple geometrical 
% interpretation (see \cite{JJSaku85,PCWKim93}). 

\section{Special Case of a Three - Layer Medium} 

\indent Similar analysis can be performed for the 
transitions of neutrinos  traversing three layers 
of constant density and chemical composition when
the first and the third layers have the same density,
chemical composition and width
which differ, however, from those of the second layer.
This corresponds to the physically 
important case of solar and atmospheric neutrinos 
crossing the Earth core
on the way to detectors located near or on the Earth surface,
the first and third layers being the Earth mantle, and
the Earth core playing the role of the second layer.

  We will use the same notations for the parameters 
of the first (third) and the second layers as in Section 2,
and will assume that eq. (\ref{poten}) holds.
The evolution matrix in the case of interest is given by:
$U = U'U''U'$. Accordingly, the parameters of $U$, 
eq. (\ref{evolution}), can 
be expressed in terms of the parameters of the
first (third) and the second layers as follows 
(see, e.g., \cite{JJSaku85}): 
\begin{equation}
\left\{
\begin{array}{l}
\cos\phi=
% \cos(2\phi')\cos\phi''-({\bf n}'\cdot{\bf n}'')\sin(2\phi')\sin\phi'',
\cos(2\phi')\cos\phi''- \cos(2\theta''_m - 2\theta'_m)\sin(2\phi')\sin\phi'',
\\
{\bf n}\sin\phi=
{\bf n}'\left[\sin(2\phi')\cos\phi''-({\bf n}'\cdot{\bf n}'')
(1-\cos(2\phi'))\sin\phi''\right] +
{\bf n}''\sin\phi''.
\end{array} \right.
\label{par-3}
\end{equation}
An expression for the probability $P_{\alpha \beta}$ 
is given in \cite{SP1} (see also \cite{s5398}).
The necessary conditions for a maximum 
of the probability $P_{\alpha \beta}$,
obtained by varying the phases $\phi'$ and $\phi''$, read
 \begin{equation}
max~P_{\alpha \beta}:~\left\{ \begin{array}{l}
\sin (2\theta_m')~ F(2\theta''_{m} - 2\theta'_{m},\pi/2; 
   \phi'',2\phi'+\pi/2) = 0,\\ 
 F(2\theta''_{m} - 2\theta'_{m},2\theta'_{m}; \phi'' + \pi/2,2\phi') = 0,
 \end{array} \right.
 \label{cond-3m}
 \end{equation}
\noindent while the requirements for an {\it absolute} maximum 
have the form
\begin{equation}
max~P_{\alpha \beta} = 1:~\left\{ \begin{array}{l}
\cos\phi \equiv 
F(2\theta''_{m} - 2\theta'_{m},\pi/2; \phi'',2\phi'+\pi/2) = 0,\\
n_3\sin\phi \equiv 
F(2\theta''_{m} - 2\theta'_{m},2\theta'_{m} + \pi/2; \phi'',2\phi') = 0.
\end{array} \right.
\label{cond-3}
\end{equation}
\noindent where the function $F(Y,Z; \varphi,\psi)$ 
is defined by eq. (\ref{F}). Using the expression
for $F(Y,Z; \varphi,\psi)$ we get the 
conditions (\ref{cond-3}) in explicit form: 
\begin{equation}
max~P_{\alpha \beta} = 1:~\left\{ \begin{array}{l}
% F(2\theta''_{m} - 2\theta'_{m},\pi/2; \phi'',2\phi'+\pi/2) = 0,\\
% \cos(2\phi')\cos\phi'' -
% \sin(2\phi')\sin(\phi'')\cos(2\theta_m''- 2\theta_m') = 0,
2\cos(\phi')\cos\Phi - \cos \phi'' = 0,
\\
% F(2\theta''_{m} - 2\theta'_{m},2\theta'_{m} + \pi/2; \phi'',2\phi') = 0,
 \sin(\phi'')\cos(2\theta''_m) + 2\sin(\phi')\cos(2\theta'_m)\cos\Phi = 0,\\ 
% + \cos 2\theta_m'[\sin\phi''\cos2\phi'\cos(2\theta_m''- 2\theta_m')
% + \cos\phi''\sin2\phi'] \\ 
% - \sin\phi''\sin 2\theta_m'\sin(2\theta_m''- 2\theta_m') = 0,
\end{array} \right.
\label{cond-3expl}
\end{equation}
\noindent where $\cos\Phi$ is given by eq. (\ref{par-2}).

The {\it absolute} maxima of $P_{\alpha \beta}$,
\begin{equation}
case~A:~max~P_{\alpha \beta} = 1,
\end{equation}

\noindent are provided by the solutions of eq. (\ref{cond-3})
(or (\ref{cond-3expl})),
which can be found explicitly:
\begin{equation}
solution~A:~\left\{ \begin{array}{l}
\tan\phi'=\pm\sqrt{{\displaystyle -\cos 2\theta_m''\over 
\displaystyle\cos(2\theta_m''- 4\theta_m')}}, \\
\tan\phi''=\pm{\displaystyle \cos 2\theta_m'\over \sqrt{
\displaystyle-\cos(2\theta_m'')\cos(2\theta_m''- 4\theta_m')}},
\end{array} \right.
\label{max-3}
\end{equation}  
\noindent where the signs are correlated.

   The probability $P_{\alpha \beta}$ 
($P_{\bar{\alpha} \bar{\beta}}$) exhibits a system of 
of maxima which is similar to that  
in the two layer case.
Under the conditions (8), (21) and if $V'_{\alpha \beta} > 0$
(i.e., for the 
$\nu_{\mu}~(\nu_e) \rightarrow \nu_e~(\nu_{\mu;\tau})$,
$\nu_e \rightarrow \nu_s$ and
$\bar{\nu}_{\mu} \rightarrow \bar{\nu}_s$ transitions in the Earth), 
solutions (\ref{max-3}) are realized in the region $A$ (Figs. 3 - 5), 
\begin{equation}
region~A:~\left\{ \begin{array}{l}
\cos(2\theta_m'') \le 0, \\ 
\cos(2\theta_m'' - 4\theta_m') \ge 0.
\end{array} \right.
\label{regionA3}
\end{equation}  
\noindent On the  line belonging 
to region $A$,
\begin{equation}
region~B:~\cos 2\theta_m' = 0,
\label{regionB3}
\end{equation}
\noindent we have 
\begin{equation}
case~B:~ max~P_{\alpha \beta} = \sin^2 2\theta_m' = 1,
\end{equation}
\noindent provided 
\begin{equation}
solution~B:~\left\{ \begin{array}{l}
\cos2\phi'=0,~or~ 2\phi' = {\pi \over {2}} (2k' + 1),~k' = 0,1,..., \\
\sin\phi''=0,~or~ 2\phi'' = 2\pi k'',~k'' = 0,1,...
\end{array} \right.
\label{B3}
\end{equation}
Besides these {\it absolute} maxima, 
there exist two regions,

\begin{equation}
region~C:~\cos(2\theta_m'') \ge 0,
\label{regionC3}
\end{equation}
\noindent and
\begin{equation}
region~D:~\cos(2\theta_m'' - 4\theta_m') \le 0,
\label{regionD3}
\end{equation}
\noindent with % {\it local} 
maxima

\begin{equation}
case~C:~max~P_{\alpha \beta} = \sin^2 2\theta_m'',
\end{equation}
\noindent and
\begin{equation}
case~D:~max~P_{\alpha \beta} = \sin^2(2\theta_m''- 4\theta_m'),
\label{maxD3}
\end{equation}

\noindent which correspond to the solutions

\begin{equation}
solution~C:~\left\{ \begin{array}{l}
\sin\phi'=0,~or~  2\phi' = 2\pi k',~k' = 0,1,...,  \\
\cos\phi''=0,~or~ 2\phi'' = \pi (2k'' + 1),~k'' = 0,1,...,
\end{array} \right.
\label{C3}
\end{equation}

\noindent and
\begin{equation}
solution~D:~\left\{ \begin{array}{l}
\cos\phi'=0,~or~ 2\phi' = \pi (2k' + 1),~k' = 0,1,..., \\  
\cos\phi''=0,~or~2\phi'' = \pi (2k'' + 1),~k'' = 0,1,...,
\end{array} \right.
\label{D3}
\end{equation}

\noindent respectively.
In contrast to the two-layer case,
the region $B$ is just a line, belonging actually to
the region A, and on it only absolute maxima 
can be realized due to the MSW effect
in the first (third) layer
(the mantle in the case of the Earth). 
The case $C$ corresponds to the MSW effect
in the second layer,
while the cases $A$ and $D$ correspond to the
new resonance-like effect of constructive 
interference between transition amplitudes
in the first (and third) and the second layers. 
Due to this effect the 
transition probability can reach
its maximal value, $P_{\alpha \beta} = 1$, 
if the oscillation parameters 
$\sin^2(2\theta)$ and $\Delta m^2/E$ 
belong to the region $A$, eq. (\ref{regionA3}).
%%%%%%%%%%%%%%%%%%%%%%%%%%%%%%%%%%%
\begin{figure}[t]
\begin{center}
\includegraphics[width=14cm,height=10cm]{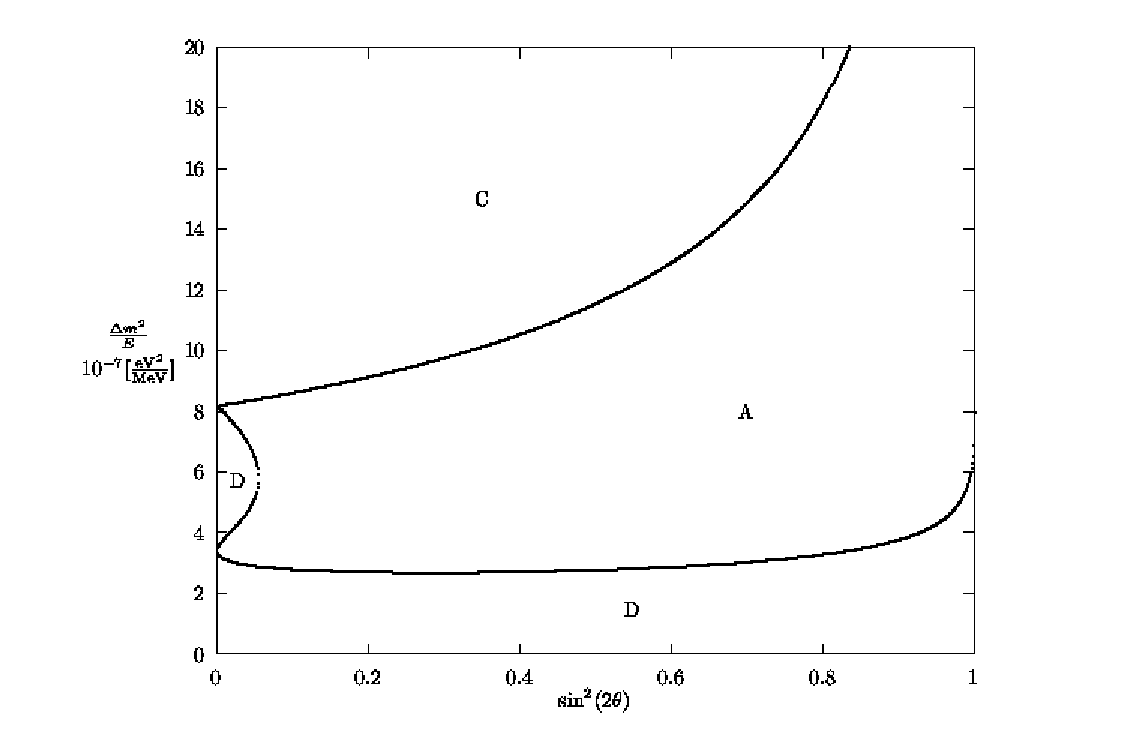}
\caption{
The regions of
the three different solutions $A$, $C$ and $D$
(eqs. (\ref{max-3}), (\ref{C3}) and 
(\ref{D3})) for the maxima of the
transition probability $P_{e\mu} = P_{\mu e}$
in a three-layer medium. The three different layers
correspond to the mantle-core-mantle of the Earth 
\protect\cite{Stacey:1977,PREM81}. 
% {\bf The figure is obtained using the average 
% (electron number) densities 
% of the Earth mantle and core.}
}
\label{Fig3}
\end{center}
\end{figure}
%%%%%%%%%%%%%%%%%%%%%%%%%%%%%%%%%%%%%%%%%
%
%%%%%%%%%%%%%%%%%%%%%%%%%%%%%%%%%%%
\begin{figure}[t]
\begin{center}
\includegraphics[width=14cm,height=10cm]{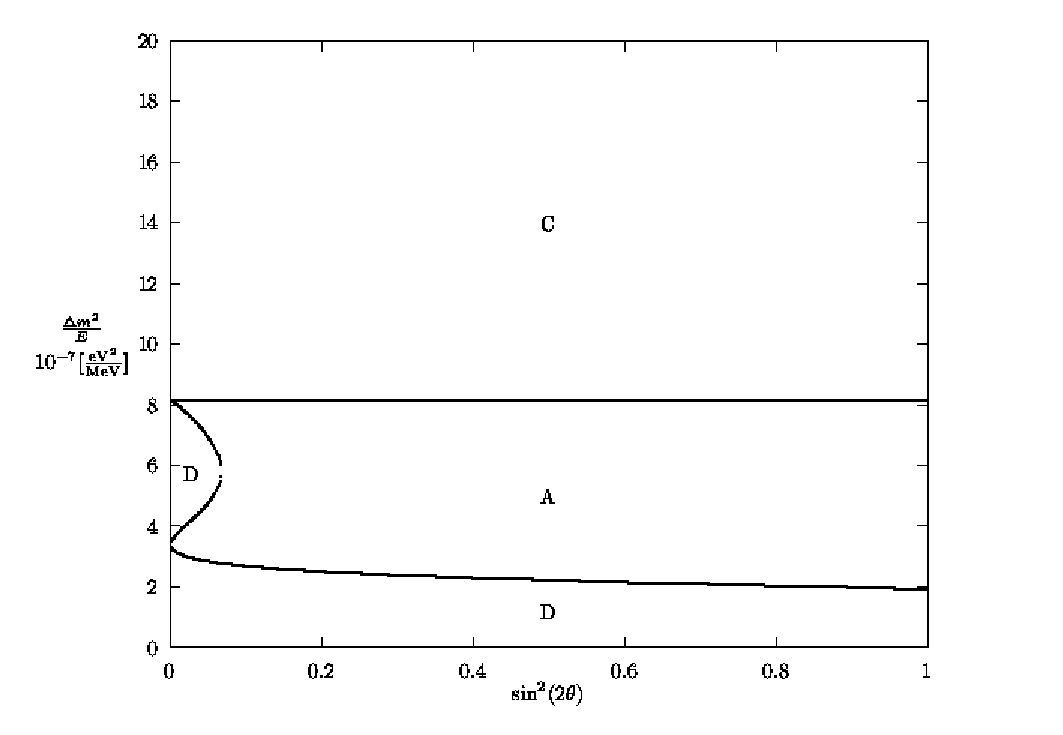}
\caption{The regions of
the three different solutions $A\equiv A^\odot$, 
$C\equiv C^\odot$ and $D\equiv D^\odot$
% (eqs. (\ref{max-Pe2}), (\ref{C3sun}) and  (\ref{D3sun})) 
(eqs. (\ref{regionAe2}), (\ref{regionC3sun}) and 
(\ref{regionD3sun}))
for the maxima of the 
probability $P_{e2}$ 
of the $\nu_2\rightarrow \nu_e$ 
transitions of neutrinos 
in a three-layer medium 
in the case of $\nu_e - \nu_{\mu (\tau)}$ mixing.
The three different layers
correspond to the mantle-core-mantle of the Earth 
\protect\cite{Stacey:1977,PREM81}.
% {\bf The figure is obtained using the average 
% (electron number) densities 
% of the Earth mantle and core.}
}
\label{Fig4}
\end{center}
\end{figure}
%%%%%%%%%%%%%%%%%%%%%%%%%%%%%%%%%%%%%%%%%
%
%%%%%%%%%%%%%%%%%%%%%%%%%%%%%%%%%%%
\begin{figure}[t]
\begin{center}
\includegraphics[width=14cm,height=10cm]{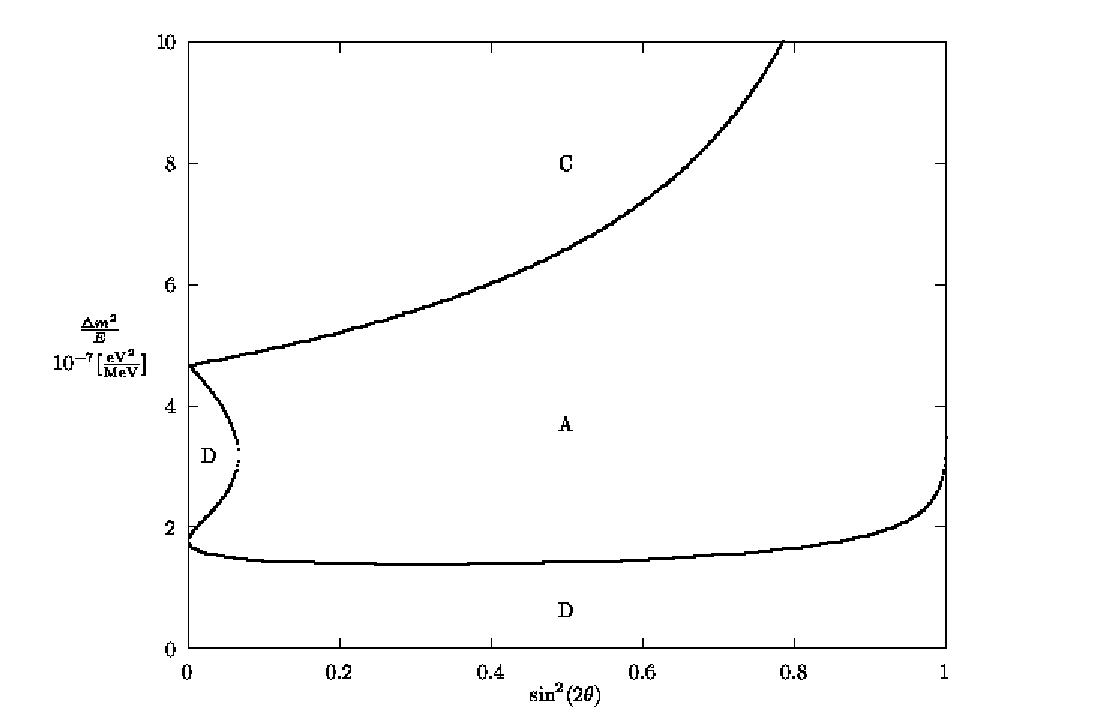}
\caption{
The same as in figure 3
for the transition probability 
$P_{\bar{\mu} \bar{s}} \equiv 
P(\bar{\nu}_{\mu} \rightarrow \bar{\nu}_s)$.
}
\label{Fig5}
\end{center}
\end{figure}
%%%%%%%%%%%%%%%%%%%%%%%%%%%%%%%%%%%%%%%%%
%

  The {\it absolute} minima of the probability  
$P_{\alpha \beta}$ % (\ref{Pem-3}) 
are determined by the equation
\begin{equation}
min~P_{\alpha \beta} = 0:
F(2\theta''_{m} - 2\theta'_{m},2\theta'_{m}; \phi'',2\phi') = 0,
% ~W(\sin(2\theta'),\sin(2\theta''))=0.
\label{min-3Pem}
\end{equation} 
\noindent while for
$F(2\theta''_{m} - 2\theta'_{m},2\theta'_{m}; \phi'',2\phi') \neq 0$,
the necessary conditions for the 
minima are given by eq. (\ref{cond-3m}).

   We get the same system of maxima also in the 
$\nu_2 \rightarrow \nu_e$ transition probability,
$P(\nu_2 \rightarrow \nu_e) \equiv P_{e2}$, 
$\nu_2$ being the heavier of the two 
mass eigenstate neutrinos in vacuum,
which can be used to account for the the Earth matter effects
in the transitions of solar neutrinos traversing the Earth.
The probability $P_{e2}$ is given 
in the case of $\nu_e \ - \nu_{\mu (\tau)}$ mixing 
by the $U_{e2}$ element, $P_{e2} = |U_{e2}|^2$, of the evolution matrix

\begin{equation}
U^\odot\equiv\left(\begin{array}{cc}
U_{e1} & U_{e2} \\
U_{\mu 1} & U_{\mu 2} \end{array}\right)=
\left(\begin{array}{cc}
U_{ee} & U_{e\mu} \\ 
U_{\mu e} & U_{\mu \mu} \end{array}\right)
\left(\begin{array}{cc}
\cos\theta & \sin\theta \\
-\sin\theta & \cos\theta \end{array}\right) 
\equiv UO(\theta),
\end{equation}

\noindent where $U = U'U''U'$. 
If $\nu_e - \nu_s$ mixing takes place, the index $\mu$ 
appearing in the above equation
has to be replaced by $s$. 
The matrix $U^\odot$ can be written in 
the general form (\ref{evolution}):
\begin{equation}   
U^\odot = U'U''U'O = 
\cos\phi^{\odot} - i(\mbox{\boldmath ${\sigma n^{\odot}}$})
\sin\phi^{\odot}.
\label{evol_sun}
\end{equation}

\noindent It is not difficult to express
the parameters of the evolution 
matrix (\ref{evol_sun}), $\phi^{\odot}$ and  
$\mbox{\boldmath ${n^{\odot}}$}$, 
in terms of the parameters
$\phi'$, $\mbox{\boldmath ${n'}$}$, $\phi''$, $\mbox{\boldmath ${n''}$}$
and $\theta$.
The transition probability $P_{e2}$ 
can be written in the form
\begin{equation}
P_{e2} = 
\left [ F(2\theta''_{m} - 2\theta'_{m},2\theta'_{m} - \theta; 
\phi'',2\phi') \right ]^2 + 
\sin^2\theta \left [F(2\theta''_{m} - 2\theta'_{m},\pi/2; 
\phi'', 2\phi' + \pi/2) \right]^2~. 
\label{Pe2}
\end{equation}
\noindent Using eqs. (\ref{Pe2}) and (\ref{F}) 
we get the necessary conditions for the {\it extrema} of 
$P_{e2}$,
\begin{equation}
 max~P_{\alpha \beta},  min~P_{\alpha \beta}:
~\left\{ 
\begin{array}{l}
 \sin (2\theta_m') F(2\theta''_{m} - 2\theta'_{m}, \pi/2; 
\phi'',2\phi' + \pi/2) = 0, \\
 F(2\theta''_{m} - 2\theta'_{m},2\theta'_{m} - \theta; \phi'' + \pi/2,2\phi') 
= 0,
 \end{array} \right.
 \label{cond-e2}
 \end{equation}
\noindent 
while from eqs. (\ref{max}) and (\ref{min}), utilizing eqs.  
(\ref{par-3}) and (\ref{evol_sun}), 
% and (\ref{Pe2}), 
one obtains the requirements for {\it absolute} extrema of
the probability $P_{e2}$: 
\begin{equation}
max~P_{e2} = 1:~\left\{\begin{array}{l}
 \cos\phi^{\odot} \equiv  F(2\theta''_{m} - 2\theta'_{m}, \pi/2; 
\phi'',2\phi' + \pi/2) = 0 \\
n_{3}^{\odot}
\sin\phi^{\odot} \equiv
F(2\theta''_{m} - 2\theta'_{m},2\theta'_{m} - \theta + \pi/2; 
\phi'',2\phi') = 0,
\end{array} \right.
\label{max-e2}
\end{equation}
\begin{equation}
min~P_{e2} = 0:~\left\{\begin{array}{l}
n_{1}^{\odot}
\sin\phi^{\odot} \equiv
F(2\theta''_{m} - 2\theta'_{m}, \pi/2; 
\phi'',2\phi' + \pi/2) = 0 \\ 
n_{2}^{\odot}
\sin\phi^{\odot} \equiv
F(2\theta''_{m} - 2\theta'_{m},2\theta'_{m} - \theta; \phi'',2\phi') = 0.
\end{array} \right.
\label{min-e2}
\end{equation}
\noindent  Conditions (\ref{max-e2}), for instance, 
have the following explicit form:
\begin{equation}
max~P_{e2} = 1:~\left\{\begin{array}{l}
2\cos(\phi')\cos\Phi - \cos \phi'' = 0,
\\
\sin(\phi'')\cos(2\theta''_m - \theta) +
2\sin(\phi')\cos(2\theta'_m - \theta)\cos\Phi  = 0.\\ 
% = \cos (2\theta_m' - \theta)[\sin\phi''\cos2\phi'
% \cos(2\theta_m''- 2\theta_m')
% + \cos\phi''\sin2\phi'] \\ 
% - \sin\phi''\sin (2\theta_m' - \theta)\sin(2\theta_m''- 2\theta_m') = 0,
\end{array} \right.
\label{cond-3explsun}
\end{equation}
\noindent These all are the same type of equations 
as (\ref{cond-3expl}) and we can easily find their solutions.

   The absolute maxima of $P_{e2}$,
\begin{equation}
case~A^\odot:max~P_{e2} = 1,
\end{equation}
\noindent are reached for
\begin{equation}
solution~A^\odot:~\left\{ \begin{array}{l}
\tan\phi'=\pm\sqrt{{\displaystyle -\cos(2\theta_m''-\theta)\over
\displaystyle\cos(2\theta_m''- 4\theta'_m + \theta)}}, \\
\tan\phi''=\pm{\displaystyle \cos(2\theta_m'-\theta)\over \sqrt{
\displaystyle-\cos(2\theta_m''-\theta)\cos(2\theta_m''- 4\theta_m'+\theta)}},
\end{array} \right.
\label{max-Pe2}
\end{equation}
\noindent where the signs are correlated,
in the region $A^\odot$ (Fig. 4),
\begin{equation}
region~A^\odot:~\left\{ \begin{array}{l}
\cos(2\theta_m''-\theta) \le 0 \\
\cos(2\theta_m''- 4\theta_m' + \theta) \ge 0.
\end{array} \right.
\label{regionAe2}
\end{equation}
\noindent There are the line
\begin{equation}
region~B^\odot:~\cos (2\theta_m' -\theta) = 0,
\label{regionB3sun}
\end{equation}
\noindent belonging to $A$, and two bordering regions,
\begin{equation}
region~C^\odot:~\cos(2\theta_m''-\theta) \ge 0,
\label{regionC3sun}
\end{equation}
\noindent and
\begin{equation}
region~D^\odot:~\cos(2\theta_m'' - 4\theta_m' + \theta) \le 0,
\label{regionD3sun}
\end{equation}
\noindent where the solutions $B$, eq. (\ref{B3}); $C$, eq. (\ref{C3});
and $D$, eq. (\ref{D3}), 
\noindent are realized.
These solutions correspond to the following maxima of $P_{e2}$:
\begin{equation}
case~B^\odot:~max~P_{e2} = \sin^2(2\theta_m'-\theta) = 1,
\end{equation}
\begin{equation}
case~C^\odot:~max~P_{e2} = \sin^2(2\theta_m''-\theta),
\label{C3sun}
\end{equation}
\noindent and
\begin{equation}
case~D^\odot:~max~P_{e2} = \sin^2(2\theta_m''- 4\theta_m' + \theta).
\label{D3sun}
\end{equation}

    Solutions $D^\odot$ and $D$ (eqs. ((\ref{regionD3}), (\ref{maxD3}),
(\ref{D3}), (\ref{regionD3sun}), (\ref{D3sun})) 
correspond to the NOLR discussed 
\footnote{Solutions $B$ and $C$ (eqs. (\ref{B3}) and (\ref{C3})) 
were considered briefly in \cite{SP1} as well.}   
in \cite{SP1}.

  For the {\it absolute} minima of $P_{e2}$ we get 
the following solutions: 
\begin{equation}
min~P_{e2} = 0:~\left\{ \begin{array}{l}
\tan\phi'=\pm\sqrt{{\displaystyle \sin(2\theta_m''-\theta)\over
\displaystyle\sin(2\theta_m'' - 4\theta_m' + \theta)}} \\
\tan\phi''=\pm{\displaystyle \sin(2\theta_m'-\theta)\over \sqrt{
\displaystyle \sin(2\theta_m''-\theta)\sin(2\theta_m'' - 4\theta_m'+\theta)}},
\end{array} \right.
\label{min-Pe2}
\end{equation}  
\noindent where again the signs are correlated.

\section{Transitions of Neutrinos Traversing the Earth Core} \indent

   In the case of transitions in the Earth of 
(solar and atmospheric) neutrinos which
pass through the Earth mantle, 
the core and the mantle again,
the two lengths $X'$ and $X''$ 
are not independent due to the 
spherical symmetry of the Earth,
if both the Earth radius and the Earth 
core radius are fixed. 
We can choose as independent variables, for example,
$\cos h$, $h$ being the Nadir angle, and $y \equiv \Delta m^2/E$.
The corresponding necessary conditions  
for the maxima of the probability 
$P_{\alpha \beta}$ read:
\begin{equation}
{{{\rm d}P_{\alpha \beta}}\over {{\rm d}\cos h}} = 0:~~~~~~~ 
{2 R_\oplus \over X''}\left(\Delta E'' R_\oplus\cos h~ 
{{\rm d}F \over {\rm d}\phi''} - % X'\Delta E'
2\phi' {{\rm d}F \over {\rm d}(2\phi')}\right) = 0,
\label{dcosh}
\end{equation}
$$
{{{\rm d}P_{\alpha \beta}}\over {{\rm d}y}} = 0:~~~~~~~
X'~{y - 2V'_{\alpha \beta}\cos(2\theta) \over 4\Delta E'}
~{{\rm d}F \over {\rm d}(2\phi')}
+\frac{X''}{2}~{y - 2V''_{\alpha \beta}
\cos(2\theta) \over
4\Delta E''}~{{\rm d}F \over {\rm d}\phi''}~~~~~ $$ 
$$ +~{V'_{\alpha \beta} \over \Delta E'^2}\sin(2\theta)\sin\phi'
\left[\cos(4\theta'_m - 2\theta''_m)
\sin\phi'\sin\phi''-\cos(2\theta'_m)\cos\phi'\cos\phi''\right]~~~~~~~~~
$$
$$-~{V''_{\alpha \beta} \over \Delta E''^2} {\sin(2\theta)\over 2}\sin\phi''
\left[\cos(4\theta'_m - 2\theta''_m)\sin^2\phi'+
\cos(2\theta''_m)\cos^2\phi' \right] = 0,~
\eqno(84)$$
\noindent where $F \equiv 
F(2\theta''_{m} - 2\theta'_{m},2\theta'_{m}; \phi'',2\phi')$ 
is given by eq. (\ref{F}),
${\rm d}F/{\rm d}\phi'' = 
F(2\theta''_{m} - 2\theta'_{m},2\theta'_{m}; \phi'' + \pi/2,2\phi')$ and
${\rm d}F/{\rm d}(2\phi') = \sin2\theta_m'
F(2\theta''_{m} - 2\theta'_{m},\pi/2; \phi'', 2\phi' + \pi/2)$.
% $y \equiv \Delta m^2/E$ and $R_\oplus$ is the radius of the Earth.
It is clear, that, in general, 
the maxima of $P_{\alpha \beta}$ in the variables 
$\phi'$ and $\phi''$
do not correspond to maxima in the variables 
$\cos h$ and $\Delta m^2/E$ (compare eqs. (\ref{cond-3m}) 
with eqs. (83) - (84)). 
It is not difficult to check, however, that
solutions $A$, eq. (\ref{max-3}), 
for the {\it absolute} maxima, 
corresponding to a total neutrino conversion,
$P_{\alpha\beta} = 1$, 
are solutions of the 
system of equations (83) - (84) as well.
They give the absolute maxima of the 
neutrino transition probability 
$P_{\alpha \beta}$ in any variable.
Equation (\ref{max-3}) gives also 
the complete set of such solutions.  
Note that solution $A$ coincides on the curves
$\cos(2\theta'_m) = 0$, 
$\cos(2\theta''_m) = 0$ and  
$\cos(2\theta''_m - 4\theta'_m) = 0$
with  $B$, $C$ and $D$, respectively. 
However, solutions $C$ and $D$, eqs. (\ref{C3}) and (\ref{D3}), 
for the {\it local} maxima with $max~(P_{\alpha\beta}) < 1$ no longer
correspond to extrema in the variables
$\cos h$ and $\Delta m^2/E$.
New solutions for the local maxima of 
$P_{\alpha\beta}$, which do not coincide 
with the solutions associated with phases
equal to multiples of $\pi$, are possible. 
The same conclusions are valid for the 
solutions $A^{\odot}$, $C^{\odot}$ and $D^{\odot}$, 
eqs. (\ref{max-Pe2}), (\ref{C3}) and 
(\ref{D3}) (giving (\ref{C3sun}), (\ref{D3sun})),
for the absolute and local 
maxima of the probability $P_{e2}$. 
Our numerical studies confirm these 
conclusions.

  The solutions $A$, eq. (\ref{max-3})
($A^{\odot}$, eq. (\ref{max-Pe2})), providing a 
total neutrino conversion, $P_{\alpha \beta} = 1$
($P_{e2} = 1$), exist for the 
Earth density profile and for all 
neutrino transitions of interest,
$\nu_2 \rightarrow \nu_{e}$,
$\nu_{\mu} \rightarrow \nu_{e}$,
$\nu_e \rightarrow \nu_{\mu(\tau)}$,
$\nu_e \rightarrow \nu_{s}$ and 
$\bar{\nu}_{\mu} \rightarrow \bar{\nu}_{s}$, and
at small, intermediate and 
large mixing angles (see Figs. 6 - 17).
They are realized for very different sets of values of 
the phases $2\phi'$ and $2\phi''$, which, 
are not multiples of $\pi$.
In Tables 2 - 4 we give a rather complete 
list of these solutions
for the transitions
$\nu_2 \rightarrow \nu_{e}$,
$\nu_{\mu} \rightarrow \nu_{e}$,
$\nu_e \rightarrow \nu_{\mu(\tau)}$,
$\nu_e \rightarrow \nu_{s}$ and 
$\bar{\nu}_{\mu} \rightarrow \bar{\nu}_{s}$ 
of the Earth-core-crossing neutrinos, while 
Figs. 6 - 17 illustrate them graphically.  
% while 
Tables 5 - 6 illustrate the behavior of
the probabilities $P_{\mu e}$ and $P_{e2}$
in the region of these solutions
at small $\sin^22\theta$ (see further).
The maximal neutrino 
conversion solutions
$A^{\odot}$ and $A$, 
eqs. (\ref{max-Pe2}) and (\ref{max-3}),  
are responsible for the 
strong resonance-like enhancement
of the $\nu_2 \rightarrow \nu_{e}$,
$\nu_{\mu} \rightarrow \nu_{e}$,
$\nu_e \rightarrow \nu_{\mu(\tau)}$,
$\nu_e \rightarrow \nu_{s}$, etc. 
transitions in the Earth of the 
Earth-core crossing solar and atmospheric 
neutrinos, discussed in \cite{SP1,s5398,SPnu98,SPNewE98}.
In \cite{SP1} this enhancement was 
interpreted to be due to the
neutrino oscillation length resonance -
solutions $D$, eq. (\ref{D3}), 
for the neutrinos passing through the Earth core. 
At small mixing angles 
the values of the parameters at which
the maximal neutrino conversion takes place
for the $\nu_2 \rightarrow \nu_{e} 
\cong \nu_{\mu} \rightarrow \nu_{e}$ and
$\nu_{e} \rightarrow \nu_{\mu}$ transitions
are rather close to the values of the parameters 
for which the NOLR giving $P_{\alpha \beta} = 1$ 
occurs (Tables 2 and 4), 
while for the 
$\nu_2 \rightarrow \nu_{e}$ transitions caused by
$\nu_e - \nu_{s}$ mixing and the
$\nu_e \rightarrow \nu_{s}$
and $\bar{\nu}_{\mu} \rightarrow \bar{\nu}_{s}$ 
transitions these values are very different
\cite{SP1,s5398} (Tables 3 - 4). 
In both cases of transitions, however,
only the maximal neutrino 
conversion mechanism is operative 
for the Earth-core-crossing neutrinos.
 
    Let us consider 
the case of $\nu_2 \rightarrow \nu_{e}$, 
$\nu_{\mu} \rightarrow \nu_{e}$ 
($\nu_{e} \rightarrow \nu_{\mu}$) 
transitions at $\sin^22\theta \ltap 0.10$ 
in somewhat greater detail.
As it follows from Table 2, for, e.g.,  
$h = 13^{\circ}$ we have   
$P_{e2} = 1$ at $\sin^22\theta = 0.044$ and 
$\Delta m^2/E = 7.1\times 10^{-7}~{\rm eV^2/MeV}$,
which corresponds to $2\phi' = 0.95\pi$ and $2\phi'' = 0.97\pi$.
% The values of $\sin^22\theta = 0.039$, 
% $\Delta m^2/E = 7.0\times 10^{-7}~{\rm eV^2/MeV}$,
% and  $2\phi' = 0.93\pi$ and $2\pi'' = 0.95\pi$
% are incorrect (and these incorrect values 
% were give in the second version sent to
% hep-ph): the correct values are those given in the text.
At the indicated point one finds also 
that $\sin^2(2\theta''_m - 4\theta'_m + \theta) = 1$,
i.e., the NOLR solution $D$ represents a very good
approximation to solution $A$. Further, 
for any fixed $\sin^22\theta < 0.044$,
% the value in the upper bound $\sin^22\theta < 0.039$,
% is incorrect and was given in the second version
% sent to the hep-ph; the correct value is given in the text.
$P_{e2}$ has an absolute maximum in the variable
$\Delta m^2/E$, which satisfies 
$max~P_{e2} < 1$
(see Figs. 6 and 7, and, e.g., Fig. 1 in \cite{SP1}, Fig.
4b in \cite{s5398}). When $\sin^22\theta$ increases from, 
say $10^{-3}$ to 0.044, 
these absolute maxima form a 
continuous curve - a ``ridge'', with respect 
to the variables $\sin^22\theta$ and $\Delta m^2/E$,
which passes through the total neutrino conversion point 
$max~P_{e2} = 1$. It was suggested in \cite{SP1} 
on the basis of several test cases studied
that the values of $P_{e2}$ forming the ``ridge'' are 
given with a relatively good precision,
by the NOLR, i.e., by eq. (3) (or (\ref{D3sun})).
These values have been calculated 
numerically and compared in Table 6 with
the one predicted by eq. (3).
We see from Table 6 
that for $\sin^22\theta = 9.0\times 10^{-3}$, 
eq. (3) reproduces the value of $max~P_{e2}$ 
of interest with an error of 32\%. 
For $h = 23^{\circ}$ and $\sin^22\theta = 10^{-2}$ 
the same error is 18\%. The error increases 
with the decreasing of
$\sin^22\theta$ and at $\sin^22\theta = 5.0\times 10^{-3}$,
for $h = 13^{\circ};~23^{\circ}$ reaches
43\%; 23\%. 

   Results of a similar analysis for the probability
$P_{e\mu} = P_{\mu e}$ are collected in Table 5 and 
are illustrated graphically in Figs. 8 - 11.
As it follows from Table 5, for $h = 0^{\circ};~23^{\circ}$
the NOLR, eq. (\ref{D3}), describes the values of $max~P_{\mu e}$ 
on the corresponding ``ridge'' 
(see Figs. 8 - 11 and Figs. 5a - 5c in \cite{s5398}) 
with an error which is 
24\%; 14\% at $\sin^22\theta = 0.01$ and 
increases to 42\%; 19\% at $\sin^22\theta = 4\times 10^{-3}$. 
More generally, Tables 5 and 6 show that
for fixed $\sin^22\theta = (10^{-3} - 10^{-2})$
and fixed $h$, the NOLR describes the absolute maxima of 
$P_{e2}$ and $P_{\mu e}$ in the variable $\Delta m^2/E$,
$max~P_{e2 (\mu e)} < 1$,
with an error which varies between 14\% and 60\%. 
%%%%%%%%%%%%%%%%%%%%%%%%%%%%%%%%%%%
\begin{figure}[t]
\begin{center}
\includegraphics[width=10cm,height=10cm]{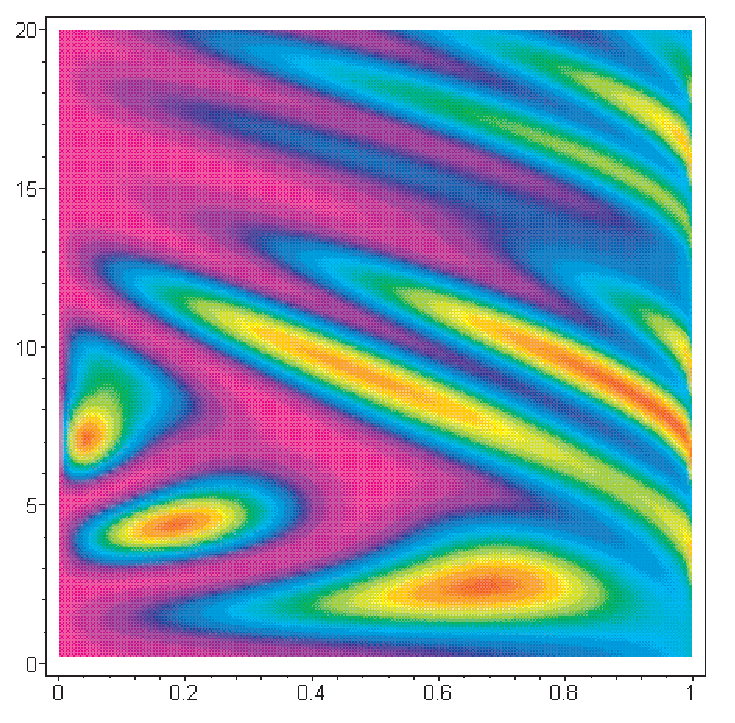}
\caption{
% The same as in figure 6 for
The probability $P_{e2}$ in the case of 
$\nu_e - \nu_{\mu (\tau)}$ mixing 
as a function of $\sin^22\theta$
(horizontal axis) and $\Delta m^2/E~[10^{-7}~{\rm eV^2/MeV}$]
(vertical axis) and for (solar) neutrinos crossing 
the Earth (core) along the trajectory
with $h = 13^{\circ}$. The ten different colors correspond to
values of $P_{e2}$ in the intervals:
0.0 - 0.1 (violet); 0.1 - 0.2 (dark blue); ...;
0.9 - 1.0 (dark red).
% and for (solar) neutrinos crossing 
% the Earth (core) along the trajectory
% with $h = 13^{\circ}$. 
The points of total neutrino
conversion (in the dark red regions), $P_{e2} = 1$,
correspond to solution % $A$
$A^{\odot}$, eq. (\ref{max-Pe2}).
}
\label{Fig610}
\end{center}
\end{figure}
%%%%%%%%%%%%%%%%%%%%%%%%%%%%%%%%%%%%%%%%
%

%%%%%%%%%%%%%%%%%%%%%%%%%%%%%%%%%%%
\begin{figure}[t]
\begin{center}
\includegraphics[width=10cm,height=10cm]{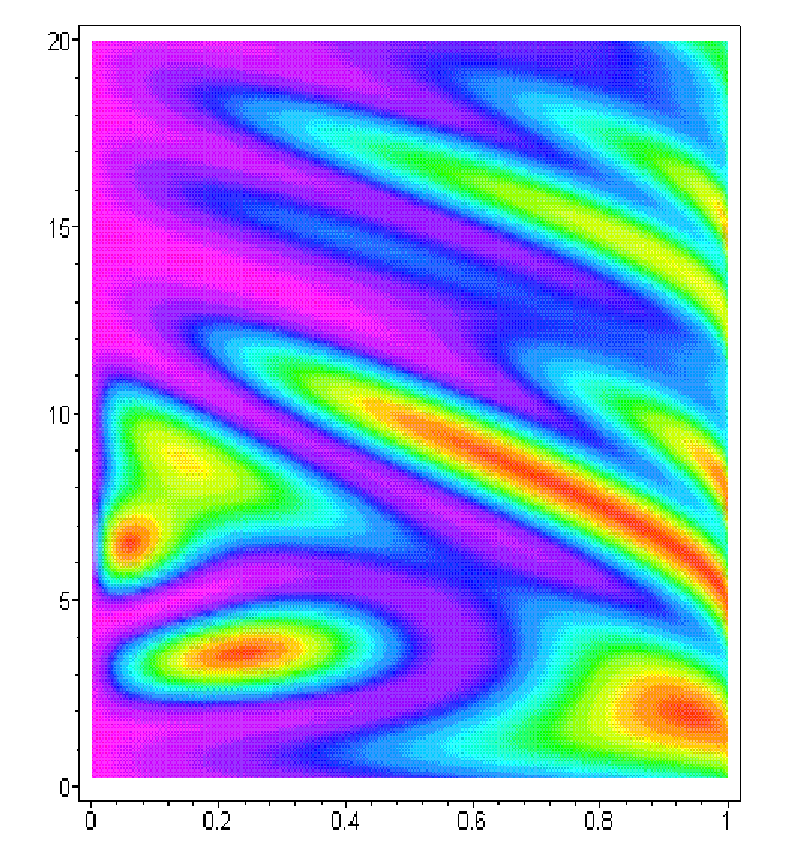}
\caption{The same as in % Fig. 10 
Fig. 6 for $h = 23^{\circ}$. 
}
\label{Fig711}
\end{center}
\end{figure}
%%%%%%%%%%%%%%%%%%%%%%%%%%%%%%%%%%%%%%%%
%
% %%%%%%%%%%%%%%%%%%%%%%%%%%%%%%%%%%%
% \begin{figure}[t]
% \begin{center}
% \includegraphics[width=10cm,height=10cm]{Fig12p3e2f23.eps}
% \caption{The same as in Fig. 10 for 
% $h = 23^{\circ}$. 
% }
% \label{Fig12}
% \end{center}
% \end{figure}
% %%%%%%%%%%%%%%%%%%%%%%%%%%%%%%%%%%%%%%%%
%

%%%%%%%%%%%%%%%%%%%%%%%%%%%%%%%%%%%
\begin{figure}[t]
\begin{center}
\includegraphics[width=10cm,height=10cm]{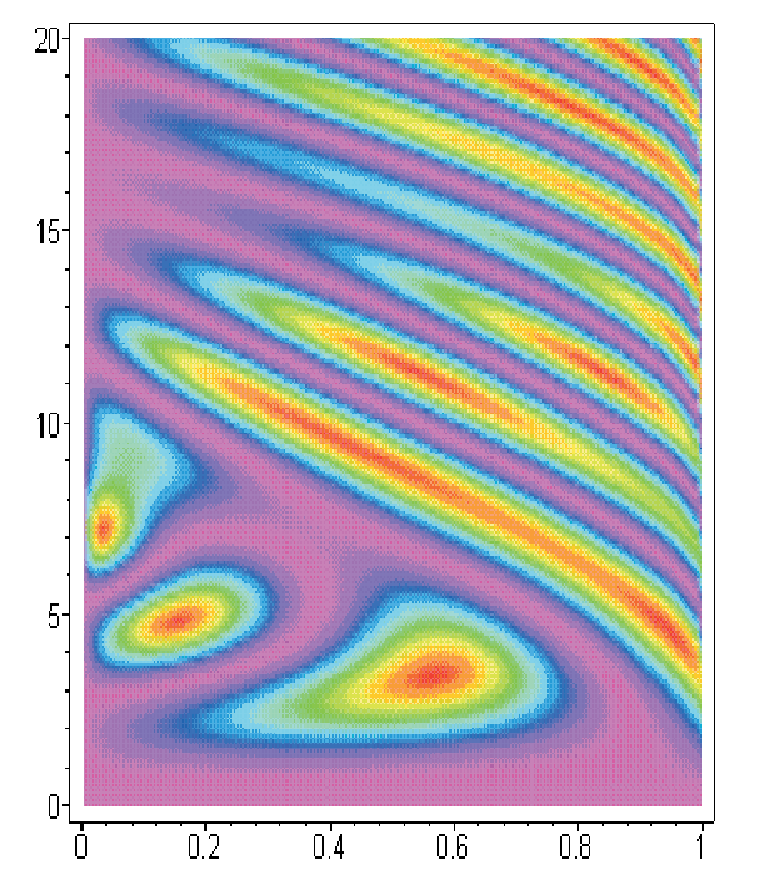}
\caption{
 The probability
$P_{e\mu} = P_{\mu e}$ for the Earth-center-crossing
(atmospheric) neutrinos 
($h = 0^{\circ}$), as a function of $\sin^22\theta$
(horizontal axis) and $\Delta m^2/E~[10^{-7}~{\rm eV^2/MeV}$]
(vertical axis). The ten different colors correspond to
values of $P_{e\mu}$ in the intervals:
0.0 - 0.1 (violet); 0.1 - 0.2 (dark blue); ...;
0.9 - 1.0 (dark red). The points of total neutrino
conversion (in the dark red regions), $P_{e\mu} = 1$,
correspond to solution $A$, eq. (\ref{max-3}),
for the Earth-core-crossing neutrinos.
}
\label{Fig86}
\end{center}
\end{figure}
%%%%%%%%%%%%%%%%%%%%%%%%%%%%%%%%%%%%%%%%
%
%%%%%%%%%%%%%%%%%%%%%%%%%%%%%%%%%%%
\begin{figure}[t]
\begin{center}
\includegraphics[width=10cm,height=10cm]{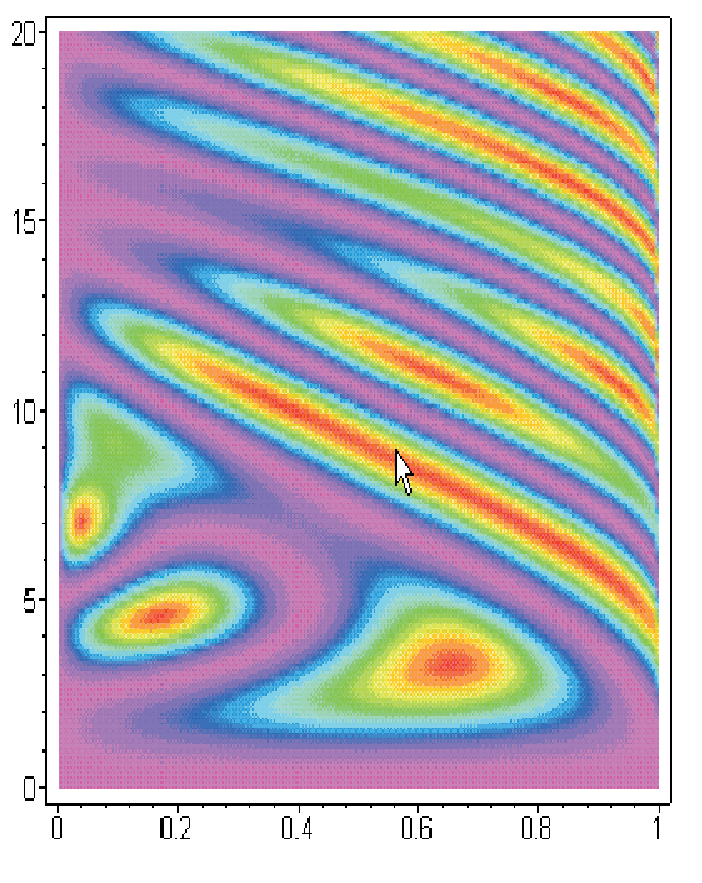}
\caption{
  The same as in Fig. 8 for $h = 13^{\circ}$. 
}
\label{Fig97}
\end{center}
\end{figure}
%%%%%%%%%%%%%%%%%%%%%%%%%%%%%%%%%%%%%%%%
%
%%%%%%%%%%%%%%%%%%%%%%%%%%%%%%%%%%%
\begin{figure}[t]
\begin{center}
\includegraphics[width=10cm,height=10cm]{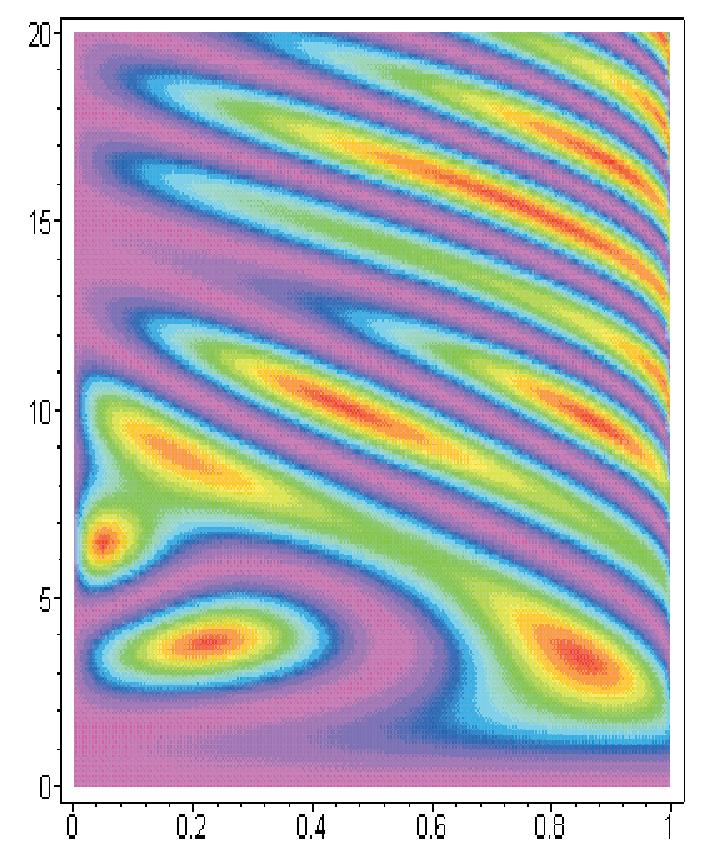}
\caption{ The same as in Fig. 8 for 
$h = 23^{\circ}$. 
 }
\label{Fig108}
\end{center}
\end{figure}
%%%%%%%%%%%%%%%%%%%%%%%%%%%%%%%%%%%%%%%%
%
%%%%%%%%%%%%%%%%%%%%%%%%%%%%%%%%%%%
\begin{figure}[t]
\begin{center}
\includegraphics[width=10cm,height=10cm]{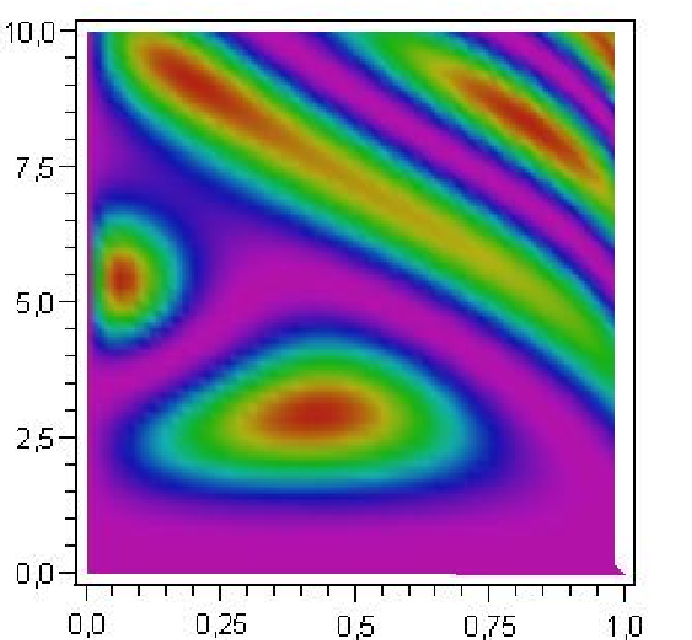}
\caption{ The same as in Fig. 8 for 
$h = 30^{\circ}$. 
}
\label{Fig119}
\end{center}
\end{figure}
%%%%%%%%%%%%%%%%%%%%%%%%%%%%%%%%%%%%%%%%
%

   In some cases the total neutrino conversion 
solutions $A$ and $A^{\circ}$ of interest
occur for values of the parameters 
which are close to those for 
the solutions $B$, eq. (\ref{B3}), 
or $C$, eq. (\ref{C3}) (Tables 2 - 4).
In all such cases the corresponding 
absolute maxima of the neutrino 
transition probabilities 
lie in the regions of the solutions 
$A$ and $A^{\circ}$, 
eqs. (\ref{regionA3}) and (\ref{regionAe2}).

     As Tables 2 - 4 and Figs. 6 - 17 indicate,
for the Earth-core-crossing neutrinos,
the new enhancement mechanism 
produces at $\sin^22\theta \ltap 0.10$
one relatively broad (in $\sin^22\theta$, 
$\Delta m^2/E$ and $h$) 
resonance-like interference peak
of total neutrino conversion 
in each of the probabilities $P_{e2}$,
$P(\nu_{\mu} \rightarrow \nu_{e})$,
$P(\nu_{e} \rightarrow \nu_{s})$ and
$P(\bar{\nu}_{\mu} \rightarrow \bar{\nu}_{s})$. 
The total conversion peaks of the different 
probabilities are located in the interval
$0.03 \ltap \sin^22\theta \ltap 0.10$.
They all take place for values of the resonance density
$N^{res} = \Delta m^2 \cos2\theta/(2E\sqrt{2}G_{F})$, which
differ from the number densities in the core and in the mantle
$V'_{\alpha \beta}/(\sqrt{2}G_{F})$ and
$V''_{\alpha \beta}/(\sqrt{2}G_{F})$, and lie between the 
latter two \cite{SP1}:
$V'_{\alpha \beta} < \sqrt{2}G_{F}N^{res}
< V''_{\alpha \beta}$.
For all the neutrino transitions considered, 
the corresponding interference peak  
is sufficiently wide
in all variables, which makes the 
transitions observable in the region 
of the enhancement 
\footnote{Let us note that the peak 
is not symmetric in any of the variables.}: 
the relative width
of the peak in  the neutrino energy
is $\Delta E/E_{max} \cong (0.3 - 0.5)$
and practically does not vary with
$\sin^22\theta$ (see also \cite{s5398,SPNewE98}), 
while in $\sin^22\theta$
it is $\sim (2.0 - 2.3)$;
the absolute width in $h$ for the peaks 
located at $h \ltap 25^{\circ}$ is
$\sim (25 - 30)^{\circ}$.
The points on the ``ridges'' leading to these peaks 
at, say,  $\sin^22\theta \ltap 0.03$
% in the $\sin^22\theta - \Delta m^2/E$ plane (Figs. 6 - 9), 
represent for fixed $h$ and $\sin^22\theta$ 
the dominating maxima in the variable $\Delta m^2/E$
of the probabilities of interest,
with $max~P_{\alpha,\beta} < 1$ 
(see Figs. 6 - 17 and , e.g., Figs. 1 - 2 in \cite{SP1} and Figs.
4b, 5a - 5c, 10a - 10b, 15a - 15c in \cite{s5398}). 
% The change of 
% $P(\nu_{\mu} \rightarrow \nu_{e}) =
% P(\nu_e \rightarrow \nu_{\mu(\tau)}) = P_{\mu e}$
% along such a ``ridge'' is illustrated by Table 5.

  As Tables 2 - 4 and Figs. 6 - 17 show, 
the absolute maxima of total neutrino conversion, 
solution $A$, are present in the probabilities $P_{e2}$,
$P(\nu_{\mu} \rightarrow \nu_{e})$,
$P(\nu_{e} \rightarrow \nu_{s})$ and
$P(\bar{\nu}_{\mu} \rightarrow \bar{\nu}_{s})$,
at large values of $\sin^22\theta \cong (0.50 - 1.0)$ as well.
They are also present in certain transitions at 
$\sin^22\theta \cong (0.15 - 0.50)$.
Although these resonance-like interference maxima 
are, in general, narrower than at small mixing 
angles (see Figs. 6 - 17), they are sufficiently 
wide and can have observable effects
both at $\sin^22\theta \cong (0.50 - 1.0)$ and
$\sin^22\theta \cong (0.15 - 0.50)$.
For certain specific trajectories the values of 
the phases $2\phi'$ and $2\phi''$ at the points of 
total conversion are close to odd multiples of
$\pi$ and the NOLR solution $D$ 
reproduces approximately the results 
corresponding to the exact solution $A$. 
This takes place
i) for $P_{\mu e}$
at $h= 23^{\circ}$ and $\sin^22\theta = 0.848$ (Table 2, Fig. 10),
ii) for $P_{e2}$ ($\nu_e - \nu_{\mu}$ mixing)
at $h= 23^{\circ}$ and $\sin^22\theta = 0.941$ (Table 4, Fig. 7),
iii) for $P_{e s}$ 
at $h= 0^{\circ}$ and $\sin^22\theta = 0.999$ and
at $h= 30^{\circ}$ and $\sin^22\theta = 0.985$ (Table 3, Fig. 15),
iv) for $P(\bar{\nu}_{\mu} \rightarrow \bar{\nu}_{s})$
at $h= 30^{\circ}$ and $\sin^22\theta = 0.981$ (Table 3).
This approximate equality of the two solutions 
is  unstable with respect
to the change of the parameters.
Inspecting the Tables 2 - 4 one finds
that for Nadir angles different from those
indicated in i) - iv),  
the total neutrino conversion (solution $A$) 
takes place for values of the parameters
at which the NOLR does not provide at all,
or provides a poor description of
the enhancement of interest.
Note also that at $\sin^22\theta \sim 1$ 
one can have $\sin^2(2\theta_m'' - 4\theta_m') \cong 1$ 
at the points of total neutrino conversion, solution $A$,
even for $2\phi'$
and $2\phi''$ 
which differ quite substantially 
from being odd multiples of $\pi$  
(as are the cases, e.g., of $max~P_{\mu e} = 1$ at 
$h = 0^{\circ}$ and $\sin^22\theta = 0.994$, as well as  
of $max~P_{es} = 1$ 
at $\sin^22\theta = 0.931$ 
and of $max~P_{\bar{\mu} \bar{s}} = 1$  at
$\sin^22\theta = 0.992$, 
for $h = 23^{\circ}$). In these cases 
the analytic expression for
$P_{\alpha \beta}$ does not coincide 
with $\sin^2(2\theta_m'' - 4\theta_m')$.
%%%%%%%%%%%%%%%%%%%%%%%%%%%%%%%%%%%
\begin{figure}[t]
\begin{center}
\includegraphics[width=10cm,height=10cm]{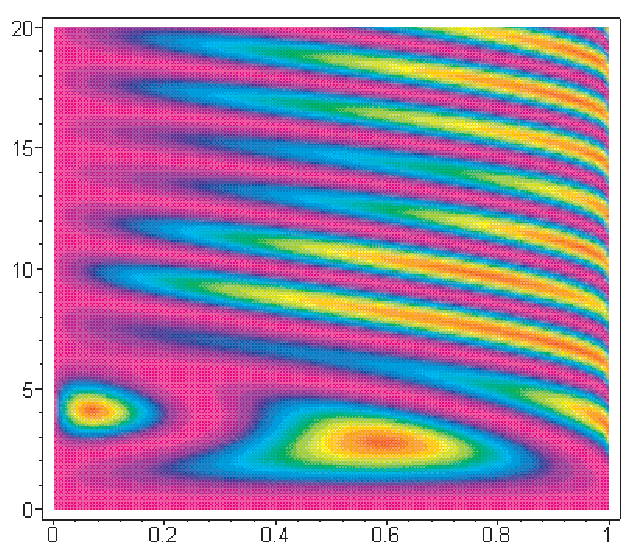}
\caption{ The same as in Fig. 8 for 
for the probability $P_{\bar{\mu} \bar{s}}$ 
and for neutrinos crossing 
the Earth (core) along the trajectory
with $h = 0^{\circ}$. The points of total neutrino
conversion (in the dark red regions), $P_{\bar{\mu} \bar{s}} = 1$,
correspond to solution $A$, eq. (\ref{max-3}).
}
\label{Fig12}
\end{center}
\end{figure}
%%%%%%%%%%%%%%%%%%%%%%%%%%%%%%%%%%%%%%%%
%
%%%%%%%%%%%%%%%%%%%%%%%%%%%%%%%%%%%
\begin{figure}[t]
\begin{center}
\includegraphics[width=10cm,height=10cm]{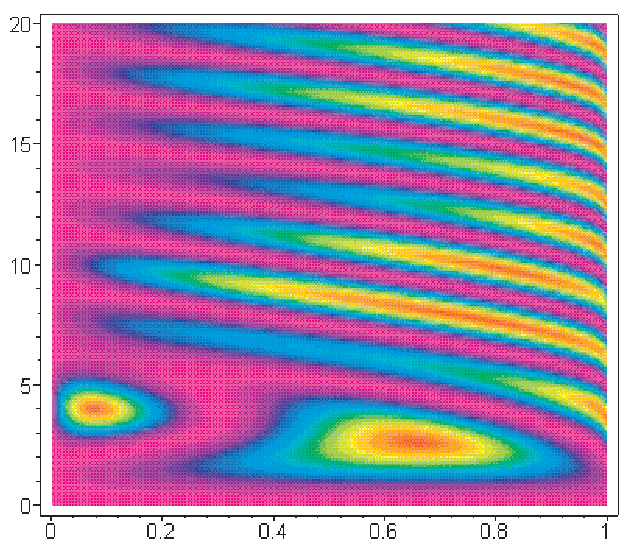}
\caption{ The same as in Fig. 12 for 
$h = 13^{\circ}$. 
}
\label{Fig13}
\end{center}
\end{figure}
%%%%%%%%%%%%%%%%%%%%%%%%%%%%%%%%%%%%%%%%
%
%%%%%%%%%%%%%%%%%%%%%%%%%%%%%%%%%%%
\begin{figure}[t]
\begin{center}
\includegraphics[width=10cm,height=10cm]{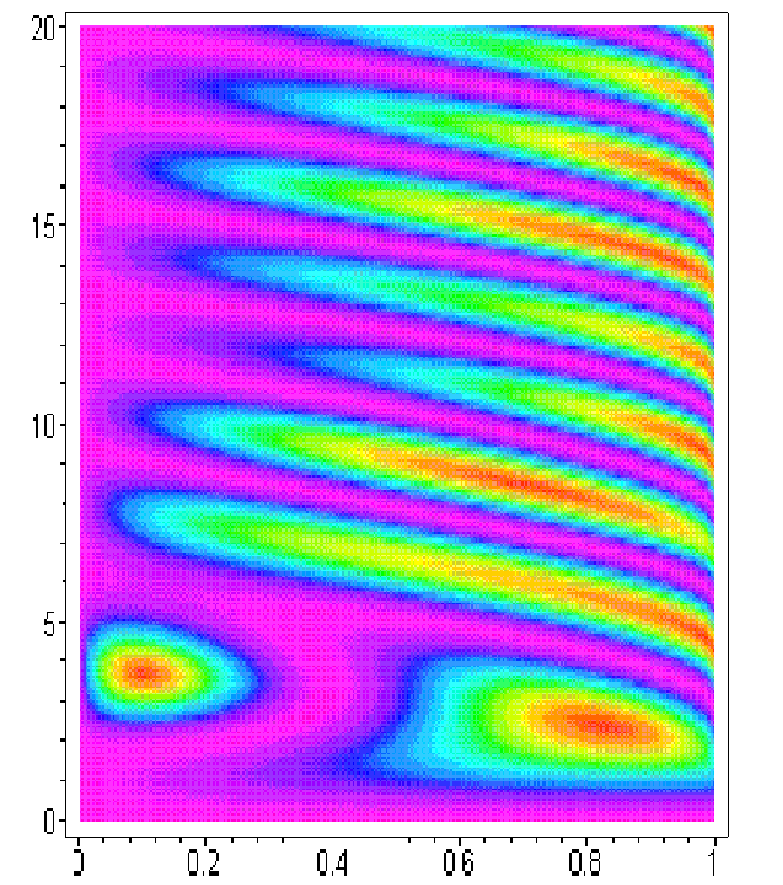}
\caption{ The same as in Fig. 12 for 
$h = 23^{\circ}$. 
}
\label{Fig14}
\end{center}
\end{figure}
%%%%%%%%%%%%%%%%%%%%%%%%%%%%%%%%%%%%%%%%
%
%%%%%%%%%%%%%%%%%%%%%%%%%%%%%%%%%%%
\begin{figure}[t]
\begin{center}
\includegraphics[width=10cm,height=10cm]{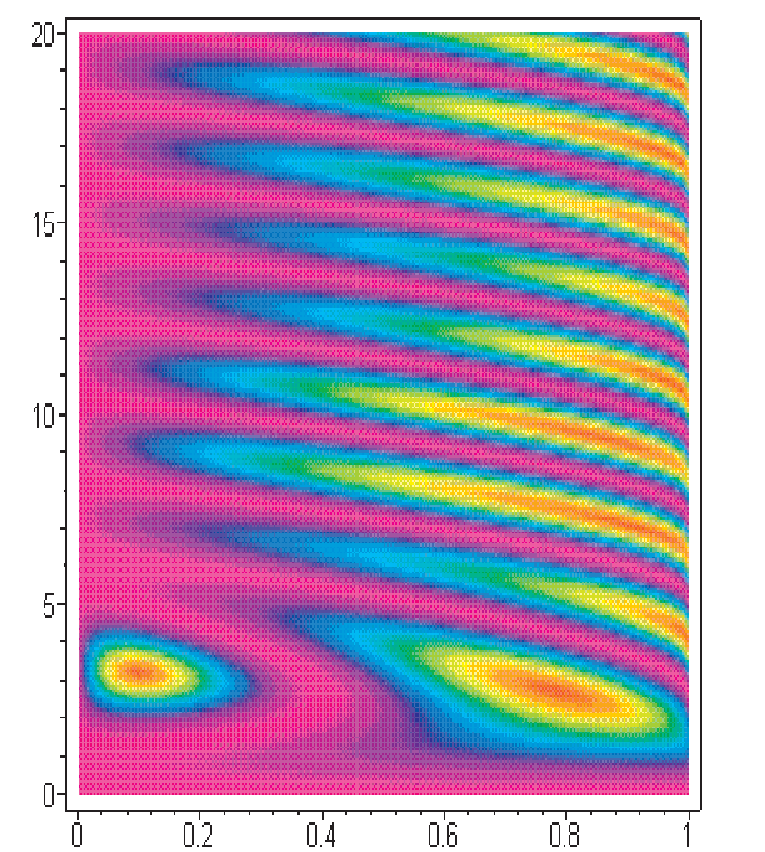}
\caption{ The same as in Fig. 8 for 
the probability $P_{e s}$
and for neutrinos crossing
the Earth (core) along the trajectory
with $h = 0^{\circ}$. The points of total neutrino
conversion (in the dark red regions), $P_{e s} = 1$,
correspond to solution $A$, eq. (\ref{max-3}).
}
\label{Fig15}
\end{center}
\end{figure}
%%%%%%%%%%%%%%%%%%%%%%%%%%%%%%%%%%%%%%%%
%
%%%%%%%%%%%%%%%%%%%%%%%%%%%%%%%%%%%
\begin{figure}[t]
\begin{center}
\includegraphics[width=10cm,height=10cm]{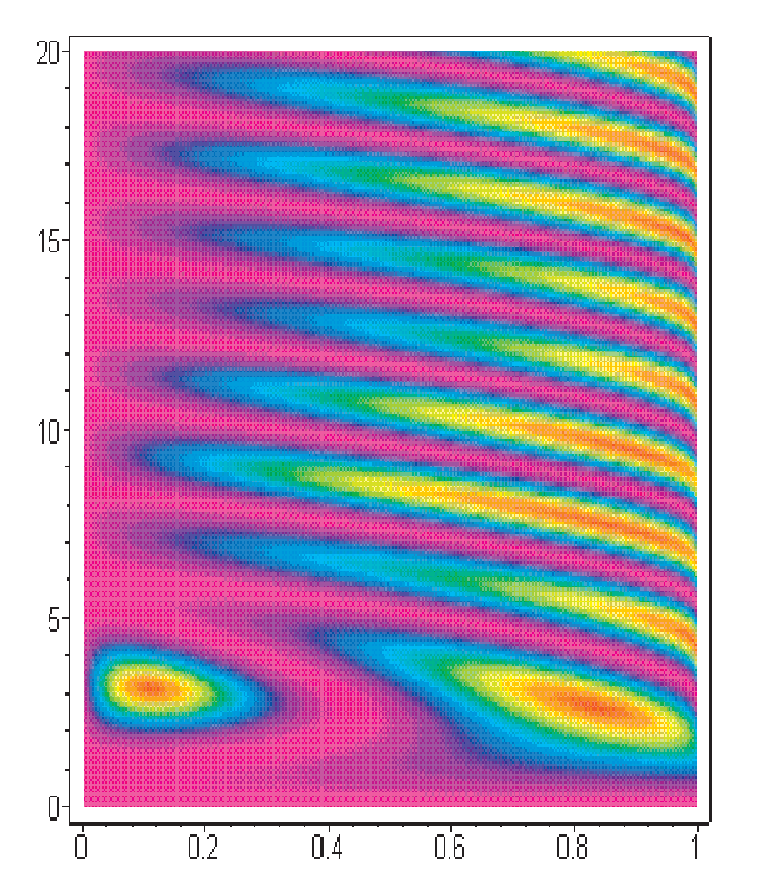}
\caption{ The same as in Fig. 15 for 
$h = 13^{\circ}$. 
}
\label{Fig16}
\end{center}
\end{figure}
%%%%%%%%%%%%%%%%%%%%%%%%%%%%%%%%%%%%%%%%
%
%%%%%%%%%%%%%%%%%%%%%%%%%%%%%%%%%%%
\begin{figure}[t]
\begin{center}
\includegraphics[width=10cm,height=10cm]{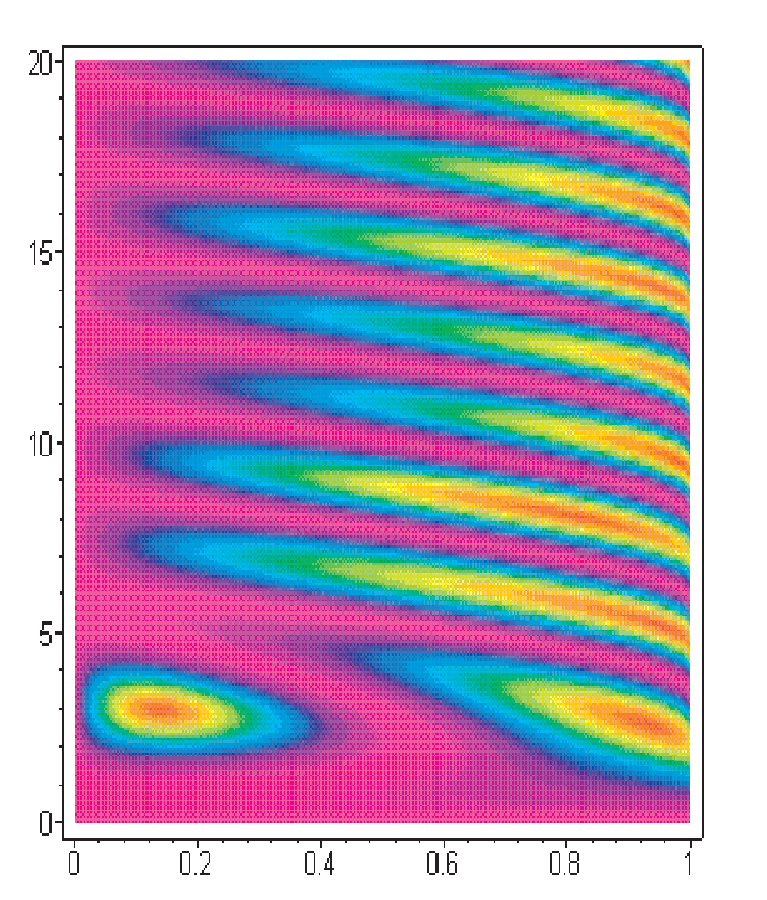}
\caption{ The same as in Fig. 15 for 
$h = 23^{\circ}$. 
}
\label{Fig17}
\end{center}
\end{figure}
%%%%%%%%%%%%%%%%%%%%%%%%%%%%%%%%%%%%%%%%
%

\subsection{Transitions of Atmospheric Neutrinos} 

\indent {\bf $\nu_{\mu} \rightarrow \nu_{e}$,
$\nu_e \rightarrow \nu_{\mu(\tau)}$.} These can be, e.g., 
sub-dominant transitions of the atmospheric $\nu_{\mu}$, 
driven by the values of $\Delta m^2$ suggested by the 
Super-Kamiokande atmospheric neutrino data \cite{SKAtmo98},

$$\Delta m^2 \cong (10^{-3} - 
8\times 10^{-3})~{\rm eV^2},~\eqno(85)$$ 
\noindent and by a
relatively small mixing \cite{SP1} (see also \cite{s5398,SPnu98,SPNewE98}). 
Such transitions should exist if
three-flavour-neutrino (or four-neutrino)
mixing takes place in vacuum, 
which is a very natural 
possibility in view of the 
present experimental evidences 
for oscillations of the flavour neutrinos.
For the Earth-center-crossing neutrinos 
there are two solutions of the type $A$, providing
a total neutrino conversion, $P_{\mu e} = P_{e\mu} = 1$,
at small mixing angles (Table 2, Figs. 8 - 11): at
$\sin^22\theta = 0.034;~0.15$ and 
$\Delta m^2/E = 7.2;~4.8\times 10^{-7}~{\rm eV^2/MeV}$. 
The first is reproduced 
approximately by 
the NOLR solution $D$
while the second is a 
new type $A$ solution. 
The above implies that for
$\Delta m^2 = 10^{-3}~{\rm eV^2}$, the 
total neutrino conversion
occurs at $E = 1.4;~2.1~{\rm GeV}$, while if 
$\Delta m^2 = 5\times 10^{-3}~{\rm eV^2}$,
it takes place at $E = 7.0;~10.5~{\rm GeV}$.
Thus, if the value of $\Delta m^2$ lies in the region (85),
the new effect of total neutrino conversion
occurs for values of the 
energy $E$ of the atmospheric
$\nu_{e}$ and  $\nu_{\mu}$ 
which contribute either to the
sub-GeV or to the multi-GeV samples of 
$e-$like and $\mu-$like events
in the Super-Kamiokande experiment.
The implications are analogous to the 
one discussed in \cite{SP1,s5398,SPnu98} in 
connection with the NOLR interpretation of the 
enhancement of $P_{\mu e}~ (P_{e\mu})$.
The new effect can produce  
an excess of e-like events
in the region $-1 \leq \cos\theta_{z}\ltap -0.8$,
$\theta_{z}$ being the Zenith angle, 
in the multi-GeV  (or a smaller one - in the
sub-GeV) sample of atmospheric neutrino events, 
and should be responsible for at
least part of the strong Zenith angle dependence, 
exhibited by the $\mu-$like multi-GeV (sub-GeV) 
Super-Kamiokande data.

  The total neutrino conversion can take 
place at several values of 
$\Delta m^2/E$ at large mixing angles (Table 2, Figs. 8 - 11),
$\sin^22\theta \gtap 0.8$. This can have implications, 
in particular, for the interpretation of the 
Super-Kamiokande data on the sub-GeV e-like 
events \cite{Giunti99,PereS99}.

\indent {\bf $\bar{\nu}_{\mu} \rightarrow \bar{\nu}_{s}$.} A total 
neutrino conversion due to
the solution $A$, eq. (\ref{max-3}), takes place 
at $\Delta m^2/E \leq 10^{-6}~{\rm eV^2/MeV}$
both at small and large mixing angles (Figs. 12 - 14). For the 
Earth-center-crossing neutrinos
the absolute maxima, $P_{\bar{\mu} \bar{s}} = 1$,
are realized at (Table 3, Fig. 12)
$\sin^22\theta = 0.07;~0.59;~0.78;~0.93;~0.999$ for 
$\Delta m^2 /E = 4.1;~2.8;~9.7;~6.7;~3.4\times 10^{-7}~{\rm eV^2/MeV}$.
For the statistically preferred value of 
$\Delta m^2 \cong 4\times 10^{-3}~{\rm eV^2}$ \cite{SKAtmo98},
this corresponds to 
$E = 9.8;~14.3;~4.12;~6.0;~11.8~{\rm GeV}$, which is 
in the range of the multi-GeV $\mu-$like events, 
studied by the Super-Kamiokande experiment.
The total neutrino conversion 
maxima are present at all  $h \ltap 30^{\circ}$, 
but at values of $\sin^22\theta$ 
and $\Delta m^2 /E$, which vary 
% quite significantly 
with $h$. This variation (and the corresponding variation of 
$2\phi'$ and $2\phi''$) is quite significant at large  
$\sin^22\theta$.  
% Note also that, e.g., for $h = 23^{\circ}$,
% $\sin^22\theta = 0.992$ and 
% $\Delta m^2/E = 4.5\times 10^{-7}~{\rm eV^2/MeV}$, 
% we have $max~P_{\bar{\mu} \bar{s}} = 1$,  
% $\sin^2(2\theta_m'' - 4\theta_m') = 0.969$,
% but $2\phi' = 1.3\pi$ and 
% and $2\phi'' = 2.4\pi$. 

 We would like to 
add the following remark in connection 
with the enhancement of the 
transitions discussed.  
In the article by
Q.Y. Liu and A. Yu. Smirnov,
Nucl. Phys. B524 (1998) 505,
it was noticed that in the case of 
muon (anti-)neutrinos crossing 
the Earth along the specific
trajectory characterized by a Nadir 
angle $h \cong 28.4^{\circ}$, and for
$\sin^22\theta \cong 1$ and
$\Delta m^2 /E \cong (1 - 2)\times 10^{-4}~{\rm eV^2/GeV}$,
the $\bar{\nu}_{\mu} \rightarrow \bar{\nu}_s$ transition
probability is enhanced
\footnote{The discussion in 
the indicated article,
being rather qualitative is done
for the $\nu_{\mu} \rightarrow \nu_s$ 
transitions as the authors
presume that $\cos2\theta \sim 0$ 
and that the matter term dominates
over the energy-dependent term
in the corresponding 
system of evolution equations.}. 
The authors interpreted the enhancement 
as being due to the conditions 
$2\phi' = \pi$ and $2\phi'' = \pi$, 
which they claimed to be approximately satisfied
and to produce {\it dominating local maxima} 
in the relevant transition probability
$P(\bar{\nu}_{\mu} \rightarrow \bar{\nu}_{s})$.
Actually, for the values of the parameters
of the examples chosen by Q.Y. Liu and A. Yu. Smirnov
to illustrate this conclusion
(Fig. 2 in their article)
one has  $2\phi' \cong (0.6 - 0.9)\pi$ and 
$2\phi'' \cong (1.2 - 1.5)\pi$. 
The indicated 
enhancement is due to the existence
of a {\it nearby total neutrino conversion point} 
which for $h \cong 28.4^{\circ}$ is located
at  $\sin^22\theta \cong 0.94$ and
$\Delta m^2 /E \cong 2.4\times 10^{-4}~{\rm eV^2/GeV}$
and at which $2\phi' \cong 0.9\pi$ and 
$2\phi'' \cong 1.1\pi$. 
This point lies in the region of solution
$A$, but very close to 
the border line with
solution $D$: the fact that the condition
$\cos(2\theta''_m - 4\theta'_m) \cong 0$ is also
fulfilled is crucial for having 
$P(\bar{\nu}_{\mu} \rightarrow \bar{\nu}_{s}) = 1$.
As we have already noticed,  
for each given $h \ltap 30^{\circ}$
there are several 
total neutrino conversion points at 
large values of $\sin^22\theta$,
at which the phases
$2\phi'$ and $2\phi''$ are not necessarily 
equal to $\pi$ or to odd multiples of $\pi$
(see Table 3).
Thus, the explanation of the enhancement
offered by the indicated authors
is at best qualitative.
% and incorrect in essence.

  \indent {\bf $\nu_{e} \rightarrow \nu_{s}$.} The atmospheric $\nu_e$ 
can undergo small mixing angle $\nu_{e} \rightarrow \nu_{s}$
transitions with $\Delta m^2$ from the interval (85).
For the Earth-core-crossing neutrinos 
the total neutrino conversion
(solution $A$) takes place 
at small mixing at $\Delta m^2/E \cong 3\times 10^{-7}~{\rm eV^2/MeV}$
(Table 3, Figs. 15 - 17), or for $E \cong (3 - 26)~{\rm GeV}$ 
if $\Delta m^2$ is given by eq. (85).
Such transitions would lead to a reduction of the rate of the
multi-GeV e-like events at $-1 \leq \cos\theta_{z}\ltap -0.8$
in the Super-Kamiokande detector.

\subsection{Transitions of Solar Neutrinos} 

\indent In the case of the transitions of the 
Earth-core-crossing solar neutrinos 
the relevant probability is $P_{e2}$. The transitions can 
be generated by $\nu_e - \nu_{\mu (\tau)}$ or by
$\nu_e - \nu_{s}$ mixing. 
The interference maxima of $P_{e2}$, 
corresponding to solution $A^{\odot}$ 
(eq. (\ref{max-Pe2})), $P_{e2} = 1$, can take place 
only in the intervals 
$\Delta m^2/E \cong (1.9 - 8.2)\times 10^{-7}~{\rm eV^2/MeV}$ 
(Fig. 4) and 
$\Delta m^2/E \cong (0.9 - 3.5)
\times 10^{-7}~{\rm eV^2/MeV}$, respectively. 

   In the case of $\nu_e - \nu_{\mu (\tau)}$ mixing, 
the absolute maximum $P_{e2} = 1$ occurs 
both at small and large mixing angles (Table 4, Figs. 6 - 7). 
At small mixing angles it takes place at 
$\sin^22\theta = 0.038;~0.044;~0.060;~0.077$ for
$h = 0^{\circ};~13^{\circ};~23^{\circ};~30^{\circ}$,
respectively. These values of $\sin^22\theta$ lie 
outside the region of the MSW SMA solution.
The change of $P_{e2}$ in the variables
$\sin^22\theta$ and $\Delta m^2/E$  along the ``ridge''
of local maxima leading to the ``peak'' $P_{e2} = 1$
at fixed $h$ is illustrated in Table 6 (see also Figs. 6 - 7, 
Figs. 1 - 2 in \cite{SP1} and Fig. 4b in \cite{s5398}). 
At $\sin^22\theta = 7\times 10^{-3}$, for instance, we have
at the ``ridge'' $P_{e2} = 0.36;~0.28$ for 
$h = 13^{\circ};~23^{\circ}$. These local maxima
dominate in $P_{e2}$: the local maxima associated with the
MSW effect in the core (mantle) are by the factor
$\sim (2.5 - 4.0)$ ($\sim (3 - 7)$) smaller \cite{SP1}.
For the values of $\Delta m^2 \cong (4 - 9)\times 10^{-6}~{\rm eV^2}$
of the MSW SMA solution region, the dominating 
local maxima take place at values of the 
solar neutrino energy from the interval
$E \cong (5.6 - 13.8)~{\rm MeV}$ \cite{SP1},
to which the Super-Kamiokande, 
SNO and ICARUS experiments are sensitive.  

  At large mixing angles, $\sin^22\theta \gtap 0.5$,
we have $P_{e2} = 1$ at values of $\sin^22\theta$ 
which fall in the region of the large mixing angle (LMA)
solution of the solar neutrino problem (Table 4, Figs. 6 - 7).
For the LMA solution values of 
$\Delta m^2 \cong (1 - 10)\times 10^{-5}~{\rm eV^2}$
the absolute maxima $P_{e2} = 1$  occur at 
$E \gtap 12.5~{\rm MeV}$ (see also \cite{s5398}) and 
typically at even larger values of the neutrino energy.
Nevertheless, their presence has an 
influence on the transitions
even at smaller $E$ and can enhance them somewhat 
(Figs. 6 - 7, see also \cite{Art2}).

   The solar $\nu_e$ can undergo SMA MSW transitions into $\nu_s$
for $\Delta m^2 \cong (3.0 - 8.0)\times 10^{-6}~{\rm eV^2}$ 
(see, e.g., \cite{KPSter96,SKDN98}).
Taking into account the interval of values of $\Delta m^2/E$ 
for which we can have total neutrino conversion 
(Table 3, Figs. 15 - 17), one finds that solution A can 
be realized only for $E \gtap 9~{\rm MeV}$.  

  The implications of 
the new enhancement effect for the interpretation of the data
of the solar neutrino experiments have been discussed in \cite{SP1}
and in much greater detail in 
\cite{Art2,Art3,MPSNODN00,MPDNCoreSK00}.
Let us just mention here that due to this enhancement 
it would be possible to probe at least part of 
the $\Delta m^2 - \sin^22\theta$ region of  the 
SMA MSW $\nu_e \rightarrow \nu_{\mu (\tau)}$ 
transition solution of the solar neutrino problem 
by performing selective D-N effect measurements 
\cite{Art2,MPSNODN00,MPDNCoreSK00}.
The Super-Kamiokande collaboration is already
exploiting successfully these results \cite{SKDN98}.  

    Recently the Super-Kamiokande 
Collaboration has published 
for the first time data on the event rate
produced only by solar neutrinos which cross the
Earth core on the way to the detector \cite{SK00}: 
in contrast to their previous night bin N5 data,
the new {\it Core} data 
are not ``contaminated'' by contributions
from neutrinos which cross the Earth mantle    
but do not cross the Earth core.
Due to the new enhancement effect,
the  Super-Kamiokande {\it Core} data and the
corresponding value of the {\it Core} D-N asymmetry
are particularly useful in performing effective 
tests the MSW SMA  
$\nu_e \rightarrow \nu_{\mu (\tau)}$
solution, as was suggested in \cite{Art2} and 
the results of  
a simplified analysis 
% of the MSW solutions,
% based on the Super-Kamiokande D-N effect data
% (including the {\it Core} one)  
show \cite{MPDNCoreSK00}.
More specifically,
at 2 s.d. a large subregion of the 
MSW SMA $\nu_e \rightarrow \nu_{\mu(\tau)}$ solution 
region is incompatible with the 
D-N effect data (including the {\it Core} one), 
while at 3 s.d. the data on the 
{\it Core} D-N asymmetry {\it alone} excludes a
non-negligible subregion of the 
indicated solution region 
(see Fig. 1c in \cite{MPDNCoreSK00}).
The results obtained in 
\cite{MPDNCoreSK00} indicate also that
the data on the {\it Core} and {\it Night}
D-N asymmetries can be used to 
perform rather effective tests of the 
MSW LMA 
% and LOW 
$\nu_e \rightarrow \nu_{\mu(\tau)}$ solution 
as well. 

  The consequences of the new enhancement 
effect for the interpretation of the solar neutrino
data are less significant
in the case of the 
MSW $\nu_e \rightarrow \nu_{s}$
solution of the solar neutrino problem 
\cite{Art3}. Nevertheless, using the  
Super-Kamiokande data on the 
{\it Core} D-N asymmetry \cite{SK00}
one can probe and constrain also the MSW 
$\nu_e \rightarrow \nu_{s}$ ``conservative'' 
solution region \cite{MPDNCoreSK00}.

    Similarly, the future data on the 
{\it Core} and {\it Night} D-N asymmetries 
in, e.g., the charged current event rate 
in the SNO detector \cite{SNO} 
can be used to perform equally
effective tests of the MSW solutions 
of the solar neutrino problem \cite{MPSNODN00}.

 \section{Conclusions} \indent

   We have investigated the extrema of 
probabilities of the
two-neutrino transitions
$\nu_{\mu}~(\nu_e) \rightarrow \nu_e~(\nu_{\mu;\tau})$,
$\nu_{2} \rightarrow \nu_e$,
$\nu_e \rightarrow \nu_s$, etc.
in a medium of nonperiodic density distribution, 
consisting of i) two layers of different 
constant densities, and ii) three
layers of constant density 
with the first and the third
layers having identical densities and widths 
which differ from those of the second layer.  
The first case would correspond, e.g., to 
neutrinos produced in the 
central region of the Earth and traversing 
both the Earth core and mantle on the way to 
the Earth surface. The second corresponds to, 
e.g., the Earth-core-crossing
solar and atmospheric neutrinos which 
pass through the Earth mantle, the core and 
the mantle again on the way to the detectors.
For both media considered we have found
that in addition 
to the local maxima corresponding 
to the MSW effect and the NOLR 
(neutrino oscillation length resonance),
there exist absolute 
maxima caused by a new effect
of enhancement and corresponding
to a total neutrino conversion, $P_{\alpha \beta} = 1$.
The conditions for existence
and the complete set of the 
new absolute maxima were derived.
These conditions differ from the
conditions of enhancement
of the probability of transitions
of neutrinos propagating in a medium with
periodically varying density,
discussed in \cite{Param}.
The absolute maxima 
associated with the new
conditions of total neutrino
conversion are absolute maxima 
corresponding to $P_{\alpha \beta} = 1$
in any independent variable characterizing 
the neutrino transitions: 
the neutrino energy,
the width of one of the layers, etc.
It was shown that the new effect of 
total neutrino conversion takes place,
in particular, in the transitions 
$\nu_{\mu}~(\nu_e) \rightarrow \nu_e~(\nu_{\mu;\tau})$,
$\nu_{2} \rightarrow \nu_e$,
$\nu_e \rightarrow \nu_s$ and
$\bar{\nu}_{\mu} \rightarrow \bar{\nu}_s$ 
in the Earth of the Earth-core-crossing 
solar and atmospheric neutrinos (Figs. 6 - 17).
The strong resonance-like 
enhancement of these transitions
discussed in \cite{SP1,s5398} 
% (see also \cite{PastAtmo,PastSun}), 
is due to the new effect. 
This enhancement was 
associated in \cite{SP1} with the NOLR 
and we show that in certain cases,
like the $\nu_2 \rightarrow \nu_e \cong
\nu_{\mu} \rightarrow \nu_e$ and 
$\nu_e \rightarrow \nu_{\mu (\tau)}$ 
transitions at small mixing angles,
the NOLR describes approximately 
the enhancement.
However, in a number of cases
the NOLR interpretation of the enhancement
fails completely. 
A ``catalog'' of the most relevant 
absolute maxima corresponding 
to a total neutrino conversion,
was given for the transitions 
indicated above.
We have shown that the NOLR and the newly found 
enhancement effect are caused by a maximal constructive
interference between the amplitudes of the neutrino
transitions in the different density layers.
Thus, the maxima they produce in 
the neutrino transition
probabilities are of interference nature. 

     Our analyses were done
under the simplest assumption
of existence of two-neutrino mixing
(with nonzero-mass neutrinos) in vacuum:
$\nu_e - \nu_{\mu (\tau)}$ or
$\nu_e - \nu_{s}$, or else
$\nu_{\mu} - \nu_{s}$.
However, in many cases of practical
interest the probabilities of the
relevant three- (or four-) neutrino mixing
transitions in the Earth reduce effectively to two-neutrino
transition probabilities for which our results are valid.
Thus, our results have a wider application than 
just in the case of two-neutrino mixing. 

  We have discussed also briefly the 
phenomenological implications of
the new effect of 
total neutrino conversion
in the transitions of the solar 
and atmospheric neutrinos
traversing the Earth core.
The effect leads to a strong 
resonance-like enhancement
of the $\nu_{2} \rightarrow \nu_{e}$
transitions of solar neutrinos
if the solar neutrino problem is due to 
MSW small mixing angle 
$\nu_{e} \rightarrow \nu_{\mu}$ 
transitions in the Sun (Figs. 6 - 7).     
The same effect should be operative also
in the $\nu_{\mu} \rightarrow \nu_{e}$
($\nu_{e} \rightarrow \nu_{\mu (\tau)}$) 
transitions of the atmospheric neutrinos crossing the Earth
core (Figs. 8 - 11) 
if the atmospheric $\nu_{\mu}$ and 
$\bar{\nu}_{\mu}$ indeed take part 
in large mixing vacuum $\nu_{\mu} \leftrightarrow \nu_{\tau}$,
$\bar{\nu}_{\mu} \leftrightarrow \bar{\nu}_{\tau}$ 
oscillations with 
$\Delta m^2 \sim (1 - 8)\times 10^{-3})~{\rm eV^2}$,
as is strongly suggested by the Super-Kamiokande
atmospheric neutrino data, and if all 
three flavour neutrinos are mixed in vacuum.
The existence of three-flavour-neutrino mixing  
in vacuum is a very natural possibility in view of the 
present experimental evidences for oscillations of
the flavour neutrinos. In both cases the 
new effect of total neutrino conversion
produces  a strong enhancement of the 
corresponding transition probabilities,
making the transitions observable 
even at rather small mixing angles (see also \cite{SP1,s5398}).
Actually, the effect may have already
manifested itself producing at least part of the
strong Zenith angle dependence in the
Super-Kamiokande multi-GeV 
$\mu-$like data \cite{SP1,s5398,SPNewE98}.
The new effect should also be present in the 
$\bar{\nu}_{\mu} \leftrightarrow \bar{\nu}_{s}$
transitions of the atmospheric 
multi-GeV  $\bar{\nu}_{\mu}$'s both at small, 
intermediate and large mixing angles, if the atmospheric 
neutrinos undergo such transitions (Figs. 12 - 13).

     As the results obtained in 
refs. \cite{SP1,s5398,Art2,MPSNODN00} 
and in the present study indicate, 
it is not excluded that some of 
the current or future high statistics solar and/or atmospheric 
neutrino experiments will be able to 
observe directly the resonance-like enhancement 
of the neutrino transitions due to the 
effect of total neutrino conversion. 
Solar neutrino 
experiments located at lower geographical latitudes 
than the existing ones or those under construction 
(see the article by J.M. Gelb et al. quoted in \cite{PastSun}) 
would be better suited
for this purpose. An atmospheric neutrino detector
capable of reconstructing with relatively good 
precision, e.g.,, the energy and 
the direction of the momentum of the 
incident neutrino in each neutrino-induced event,
would allow to study in detail the 
new effect of total neutrino conversion 
in the oscillations of 
the atmospheric neutrinos crossing the Earth.

\vskip 0.3cm
\noindent {\bf Acknowledgements.} This work
was supported in part by the Italian MURST
under the program ``Fisica Teorica delle
Interazioni Fondamentali''  and by Grant PH-510 from the
Bulgarian Science Foundation.

\newpage
% \pagebreak[1]

\newpage

\renewcommand{\arraystretch}{0.6}  % {\arraystretch}{0.5}
\begin{table} % \label{tab:tab2a}
% Subject: 2 LAYERS (Earth Core and Mantle)
\noindent \caption{
The values of 
$\Delta m^2/E \leq 20\times10^{-7}~{\rm eV^2/MeV}$ 
and of $\sin^22\theta$, at which the absolute maxima
of $P_{e\mu} = P_{\mu e}$, $P_{es}$ and $P_{\bar{\mu} \bar{s}}$, 
corresponding to a total neutrino conversion,
$P_{\alpha \beta} = 1$ (solutions $A$, eq. (\ref{max-2As})), 
take place for neutrinos 
traveling from the center of the Earth
to the Earth surface.
The values of 
$2\phi'$, $2\phi''$ (in units of $\pi$), $\sin^22\theta'_m$,
$\sin^22\theta''_m$ and $\sin^2(2\theta''_m - 2\theta'_m)$
at the absolute maxima are also given.
}
\vskip -0.6cm
\begin{center}
\begin{tabular}{|cc||cc||ccc|}
\hline
\hline
\multicolumn{7}{|c|}{$P(\nu_{\mu} \rightarrow \nu_e) = 
P(\nu_e \rightarrow \nu_{\mu (\tau)})$}\\
\hline  
\hline

 ${{\Delta m^2}\over E}~[10^{-7}~{{eV^2}\over{MeV}}]$  & 
     $\sin^22\theta$ & $2\phi'/\pi$ & $2\phi''/\pi$ &  
$\sin^22\theta'_m$ & $\sin^22\theta''_{m}$ & 
$\sin^2(2\theta''_m - 2\theta'_m)$ \\

\hline

 6.611 &    .138 &    .867  &   .888 &     .434 &    .603  &   .999 \\
 5.825 &    .651  &  1.094   &  1.866 &    .999 &    .501  &   .521 \\
11.736 &    .649  &  2.356   &  2.679 &    .872 &    .985  &   .221 \\
18.970 &    .828  &  4.153   &  4.853 &    .935 &   1.000  &   .072 \\
13.589 &    .938  &  3.063   &  3.933 &   1.000 &    .885  &   .117 \\

\hline
\hline
\multicolumn{7}{|c|}{$P(\nu_{e} \rightarrow \nu_{s})$}\\
\hline  
\hline

 3.355 &    .359 &    .531 &    .608 &    .774 &    .862  &   .590 \\
 6.073 &    .883 &   1.333 &   1.652 &    .993 &    .943  &   .103 \\
10.181 &    .904 &   2.281 &   2.723 &    .975 &    .999  &   .037 \\
13.663 &    .984 &   3.155 &   3.844 &   1.000 &    .983  &   .019 \\
17.809 &    .962 &   4.089 &   4.912 &    .989 &   1.000  &   .012 \\

\hline
\hline
\multicolumn{7}{|c|}{
$P(\bar{\nu}_{\mu} \rightarrow \bar{\nu}_{s})$}\\
\hline  
\hline

 4.006 &    .275 &    .625 &    .684 &    .612 &    .745  &   .870 \\
 5.966 &    .850 &   1.287 &   1.679 &    .989 &    .848  &   .234 \\
10.441 &    .849 &   2.303 &   2.709 &    .945 &    .996  &   .086 \\
13.589 &    .981 &   3.133 &   3.864 &   1.000 &    .959  &   .045 \\
18.002 &    .936 &   4.106 &   4.896 &    .974 &   1.000  &   .027 \\

\hline
\hline

\end{tabular}
\end{center}
\end{table}
~\vfill
\vskip -1.5cm

\renewcommand{\arraystretch}{0.6}  % {\arraystretch}{0.5}
\begin{table} % \label{tab:tab2a}
\noindent \caption{
The values of 
$\Delta m^2/E \leq 20\times10^{-7}~{\rm eV^2/MeV}$ 
and of $\sin^22\theta$, at which the absolute maxima
of $P_{e\mu} = P_{\mu e}$,
corresponding to a total neutrino conversion,
$P_{e\mu} = 1$ (solution $A$, eq. (\ref{max-3})), 
take place for neutrinos crossing
the Earth along trajectories 
passing through the Earth core:
$h = 0^{\circ};~13^{\circ};~23^{\circ};~30^{\circ}$. 
The values of 
$2\phi'$, $2\phi''$ (in units of $\pi$), $\sin^22\theta'_m$,
$\sin^22\theta''_m$ and $\sin^2(2\theta''_m - 4\theta'_m)$
at the absolute maxima are also given.
% 3 LAYERS: Earth Mantle, Core, Mantle
}
\vskip -0.6cm
\begin{center}
\begin{tabular}{|c||cc||cc||ccc|}
\hline
\hline
\multicolumn{8}{|c|}{$P(\nu_e \rightarrow \nu_{\mu (\tau)}) = 
P(\nu_{\mu} \rightarrow \nu_{e})$}\\
\hline  
\hline
h$^{\circ}$ & ${{\Delta m^2}\over E}~[10^{-7}~{{eV^2}\over{MeV}}]$  & 
     $\sin^22\theta$ & $2\phi'/\pi$ & $2\phi''/\pi$ &  
$\sin^22\theta'_m$ & $\sin^22\theta''_{m}$ & 
$\sin^2(2\theta''_m - 4\theta'_m)$ \\

\hline

 0. &   7.201 &    .034  &   .923  &   .949 &    .111 &     .614 &   1.000 \\
  0. &    4.821 &    .154 &    .509  &  2.334  &   .749  &    .208  &   .296 \\
  0. &  11.256  &   .547  &  2.174   & 4.693   &  .795   &  .996    & .285   \\
  0. &   3.407  &   .568  &   .65   & 3.612   &  .846   &  .160    & .881  \\
  0. &  18.113  &   .803  &  3.933   & 9.129   &  .923   & 1.000    & .710  \\
  0. &  11.292  &   .841  &  2.425   & 6.167   &  .987   &  .891    & .716  \\
  0. &  19.307  &   .903  &  4.314   & 10.382  &   .979  &   .987   &  .845 \\
  0. &   4.587  &   .948  &  1.167   & 4.705   &  .792   &  .285    & .996  \\
  0. &  19.862  &   .994  &  4.628   & 11.738  &   .992  &   .900   &  .981  \\
  0. &  11.373  &   .994  &  2.702   & 7.577   &  .954   &  .708    & .981  \\

\hline 

 13. &    7.024 &    .039 &    .931  &   .950  &   .131  &   .553 &   1.000 \\
 13. &   4.537  &   .169  &   .496   &  2.258  &   .845  &   .179 &    .133 \\
 13. &  10.071  &   .379  &  1.887   & 3.180   &  .645   &  .999  &   .068 \\
 13. &  10.954  &   .626  &  2.272   & 4.503   &  .869   &  .973  &   .383  \\
 13. &   3.267  &   .654  &   .732   & 3.456   &  .777   &  .153  &   .966  \\
 13. &   6.736  &   .802  &  1.477   & 4.059   &  .997   &  .580  &   .688  \\
 13. &  15.802  &   .874  &  3.653   & 7.680   &  .977   &  .972  &   .791 \\
 13. &  18.249  &   .843  &  4.205   & 8.603   &  .949   &  .997  &   .762  \\
 13. &  10.741  &   .927  &  2.529   & 5.936   &  .998   &  .797   &  .859  \\
 13. &  18.770  &   .979  &  4.542   & 9.914   &  .999   &  .922  &   .954  \\

\hline

 23. &    6.460 &    .051 &    .910  &   .922  &   .195  &   .387  &  1.000 \\
 23. &   3.812  &   .224  &   .499   & 2.008   & 1.000   &  .125   &  .111 \\
 23. &  10.101  &   .449  &  2.194   & 2.675   &  .726   &  .991   &  .134  \\
 23. &  15.419  &   .744  &  3.875   & 5.236   &  .899   &  .999   &  .612  \\
 23. &   3.417  &   .848  &  1.031   & 2.948   &  .711   &  .177   &  1.000 \\
 23. &   9.655  &  .875   & 2.495    & 4.011   & 1.000   &  .786   &  .772  \\
 23. &  16.565  &   .897  &  4.368   & 6.275   &  .984   &  .969   &  .828  \\
 23. &  20.010  &   .979  &  5.470   & 8.061   &  .999   &  .935   &  .955  \\

\hline

 30. &    5.376 &    .065 &    .765  &   .738   &  .352   &  .180  &   .982 \\
 30. &   9.038  &   .199  &  2.062   &  .920    & .418    & 1.000  &   .030  \\
 30. &   2.908  &   .437  &   .742   & 1.427    & .735    & .095   &  .969   \\
 30. &  14.060  &   .707  &  4.159   & 2.702    & .885    & .998   &  .548   \\
 30. &   8.290  &   .825  &  2.491   & 2.021    & 1.000   &  .723  &   .691  \\
 30. &  19.839  &   .913  &  6.326   & 4.357    & .982    &  .986  &   .860  \\
 30. &  17.123  &   .952  &  5.528   & 3.941    & .999    & .936   &  .909   \\
 30. &  13.989  &   .997  &  4.694   & 3.597    & .969    & .785   &  .984  \\
 30. &   8.909  &  1.000  &  3.139   & 2.741    & .882    & .551   &  .999  \\
\hline
\hline

\end{tabular}
\end{center}
\end{table}
~\vfill
\vskip -1.5cm

% \newpage
\renewcommand{\arraystretch}{0.6}  % {\arraystretch}{0.5}
\begin{table} % \label{tab:tab2a}
\noindent \caption{The same as in Table 2 for the probabilities
$P_{es}$ and $P_{\bar{\mu} \bar{s}}$ and for 
$\Delta m^2/E \leq 10\times 10^{-7}~{\rm eV^2/MeV}$.
%  3 LAYERS: Earth Mantle, Core, Mantle
}
\vskip -0.7truecm
\begin{center}
\begin{tabular}{|c||cc||cc||ccc|}
\hline
\hline
\multicolumn{8}{|c|}{$P(\nu_{e} \rightarrow \nu_{s})$}\\
\hline  
\hline
h$^{\circ}$ & ${{\Delta m^2}\over E}~[10^{-7}~{{eV^2}\over{MeV}}]$  & 
     $\sin^22\theta$ & $2\phi'/\pi$ & $2\phi''/\pi$ &  
$\sin^22\theta'_m$ & $\sin^22\theta''_{m}$ & 
$\sin^2(2\theta''_m - 4\theta'_m)$ \\
\hline

  0. &    3.202 &    .101 &    .409  &   .626  &   .334  &   .834  &   .473 \\
  0. &   2.715  &   .784  &   .565   & 1.843   &  .981   &  .538   &  .797  \\
  0. &   9.334  &   .863  &  2.064   & 4.876   &  .956   & 1.000   &  .830  \\
  0. &   6.997  &   .920  &  1.564   & 3.867   &  .997   &  .952   &  .892  \\
  0. &   4.101  &   .999  &  1.013   & 2.976   &  .887   &  .599   & 1.000  \\

\hline

 13. &   3.144  &   .110  &   .420  &   .599   &  .370   &  .799  &   .444  \\
 13. &   2.660  &   .833  &   .606  &  1.751   &  .957   &  .505  &   .873  \\
 13. &   7.456  &   .859  &  1.714  &  3.560   &  .971   &  .990  &   .816  \\
 13. &   9.450  &   .893  &  2.213  &  4.581   &  .973   &  .998  &   .865  \\
 13. &   4.324  &   .998  &  1.111  &  2.789   &  .903   &  .630  &  1.000  \\

\hline

 23. &   2.965  &   .139  &   .441  &   .523   &  .480   &  .691 &    .346 \\
 23. &   8.045  &   .817  &  2.072  &  2.861   &  .940   & 1.000 &    .768 \\
 23. &   2.605  &   .931  &   .739  &  1.482   &  .883   &  .445 &    .980 \\
 23. &   5.160  &   .958  &  1.403  &  2.206   &  .989   &  .812 &    .946  \\
 23. &   9.433  &   .987  &  2.593  &  3.810   &  .997   &  .937 &    .981  \\

\hline

 30. &   2.611  &   .196  &   .454  &   .372  &   .710 &    .503  &   .117 \\
 30. &   6.358  &   .720  &  1.878  &  1.232  &   .903 &    .999  &   .628 \\ 
 30. &   7.854  &   .978  &  2.573  &  1.850  &   .997 &    .919  &   .968  \\
 30. &   2.835  &   .985  &  1.020  &   .964  &   .834 &    .444  &  1.000  \\
 30. &   9.994  &   .996  &  3.317  &  2.371  &   .990 &    .923  &   .993  \\

\hline
\hline
\multicolumn{8}{|c|}{$P(\bar{\nu}_{\mu} \rightarrow \bar{\nu}_{s})$}\\
\hline  
\hline

  0. &   4.098 &    .070  &   .574  &   .719  &   .192  &   .713  &   .887 \\
  0. &   2.771 &    .593  &   .497  &  2.011  &  1.000  &   .356  &   .329 \\
  0. &   9.686 &    .782  &  2.096  &  4.817  &   .905  &  1.000  &   .641  \\
  0. &   6.748 &    .926  &  1.511  &  3.976  &  1.000  &   .843  &   .820  \\
  0. &   3.428 &    .999  &   .886  &  3.205  &   .811  &   .361  &   .999  \\

\hline

 13. &   4.000  &   .077  &   .582   &  .701   &  .218  &   .662  &   .881  \\
 13. &   2.635  &   .656  &   .524   & 1.926   &  .992  &   .323  &   .497  \\
 13. &   7.750  &   .768  &  1.730   & 3.511   &  .920  &   .983  &   .581  \\
 13. &   9.698  &   .834  &  2.231   & 4.554   &  .941  &   .994  &   .711  \\

\hline

 23. &    3.700 &    .101 &    .587  &   .642  &   .306  &   .520  &   .842  \\
 23. &   8.415  &   .715  &  2.107   & 2.800   &  .869   & 1.000   &  .525  \\
 23. &   2.382  &   .825  &   .632   & 1.667   &  .895   &  .261   &  .865   \\
 23. &   9.327  &   .983  &  2.558   & 3.879   &  .997   &  .880   &  .942  \\
 23. &   4.528  &   .992  &  1.299   & 2.436   &  .919   &  .531   &  .969  \\

\hline

 30. &   3.114 &    .150  &   .548   &  .488   &  .530    & .317    & .625  \\
 30. &   6.798 &    .551  &  1.913   & 1.152   &  .762    & 1.000   &  .260 \\
 30. &   7.795 &    .970  &  2.544   & 1.907   &  .997    & .845    & .911  \\
 30. &   2.450 &    .981  &   .927   & 1.127   &  .751    & .242    & 1.000 \\
 30. &   9.742 &   1.000  &  3.274   & 2.463   &  .969    & .815    & .992  \\

\hline
\hline
\end{tabular}
\end{center}
\end{table}

% \newpage
\vskip -3.0cm
\renewcommand{\arraystretch}{0.6}  % {\arraystretch}{0.5}
\begin{table} % \label{tab:tab2a}
\noindent \caption{The same as in Table 2 for the probability
$P_{e2}$ (solution $A^{\odot}$, eq. (\ref{max-Pe2}))
in the cases of $\nu_{e}  
- \nu_{\mu (\tau)}$ and
$\nu_{e}  - \nu_{s}$  
mixing, and for 
$\Delta m^2/E \leq 10\times 10^{-7}~{\rm eV^2/MeV}$.
% % 3 LAYERS: Earth Mantle, Core, Mantle; 
% % Subject: P_{2e}
}
\vskip 0.4truecm
% \begin{center}
\begin{tabular}{|c||cc||cc||ccc|}
\hline
\hline
\multicolumn{8}{|c|}{$P(\nu_{2} \rightarrow \nu_{e})$ 
($\nu_{e} % \rightarrow 
- \nu_{\mu (\tau)}$ mixing)}\\
\hline  
\hline
h$^{\circ}$ & ${{\Delta m^2}\over E}~[10^{-7}~{{eV^2}\over{MeV}}]$  & 
     $\sin^22\theta$ & $2\phi'/\pi$ & $2\phi''/\pi$ &  
$\sin^22\theta'_m$ & $\sin^22\theta''_{m}$ & 
$\sin^2(2\theta''_m - 4\theta'_m + \theta)$ \\
\hline

  0. &   7.275 &    .038  &   .944  &   .971 &    .122  &   .671 &   1.000 \\
  0. &   4.684 &    .168  &   .498  &  2.423 &    .809  &   .199 &    .370 \\
  0. &   2.724 &    .574  &   .602  &  3.755 &    .637  &   .096 &    .854 \\

\hline

 13. &   7.100  &   .044  &   .953  &   .974  &   .145  &   .611  &  1.000 \\
 13. &   4.370  &   .183  &   .480  &  2.345  &   .902  &   .167  &   .178 \\
 13. &   2.434  &   .671  &   .680  &  3.599  &   .513  &   .081  &   .953  \\
 13. &   8.025  &   .943  &  1.937  &  5.111  &   .967  &   .611  &   .314  \\

\hline

 23. &   6.507  &   .060  &   .931  &   .951  &   .223  &   .434  &  1.000 \\
 23. &   3.558  &   .235  &   .480  &  2.088  &   .983  &   .106  &   .110 \\
 23. &   7.915  &   .751  &  1.902  &  3.160  &   .992  &   .730  &   .099 \\
 23. &   5.939  &   .956  &  1.705  &  3.545  &   .885  &   .416  &   .428  \\
 23. &   1.920  &   .941  &   .945  &  3.103  &   .296  &   .056  &  1.000  \\

\hline

 30. &   5.314  &   .077  &   .761  &   .767  &   .408  &   .192  &   .984 \\
 30. &   7.800  &   .385  &  1.847  &  1.196  &   .752  &   .853  &   .040 \\
 30. &   2.425  &   .437  &   .730  &  1.486  &   .528  &   .061  &   .972 \\
 30. &   7.183  &   .489  &  1.765  &  1.334  &   .886  &   .738  &   .002 \\
\hline
\hline

\multicolumn{8}{|c|}{$P(\nu_{2} \rightarrow \nu_{e})$ 
($\nu_{e} 
- \nu_{s}$ mixing)}\\ %P_{2e} sterile

 \hline
 \hline
  0.  &  3.086  &   .122  &   .391  &   .692 &    .410  &   .765 &    .581 \\
  0. &   1.646  &   .973  &   .488  &  2.026 &    .599  &   .203 &    .823 \\
 \hline

 13. &   3.017  &   .134  &   .400  &   .666  &   .457  &   .724 &    .542  \\
 13. &   1.493  &  1.000  &   .531  &  1.934  &   .472  &   .156 &    .904 \\

 \hline

 23. &   2.802  &   .173  &   .415   &  .589   &  .601  &   .605  &   .405 \\

\hline

  30. &   2.373  &   .251  &   .421   &  .425    & .873  &   .407   &  .087 \\
 
\hline
\hline
\end{tabular}
% \end{center}
\end{table}

\vskip -3.0cm
\renewcommand{\arraystretch}{0.6}  % {\arraystretch}{0.5}
\begin{table} % \label{tab:tab2a}
\noindent \caption{Values of $P_{e \mu} = P_{\mu e}$ 
along the ``ridge'' of local
maxima in the variables
$\Delta m^2/E$ (in units of $10^{-7}$ eV$^2$/MeV) 
and $\sin^22\theta \leq 0.05$ for fixed $h$, when 
$\sin^22\theta$ decreases from the value
at which the absolute maximum of
$P_{e \mu} = 1$ (i.e., total neutrino conversion)
takes place. The results correspond 
to Earth-core-crossing neutrinos:
$h = 0^{\circ};~13^{\circ};~23^{\circ}$. 
The values of $\sin^22\theta'_m$,
$\sin^22\theta''_m$, $\sin^2(2\theta''_m - 4\theta'_m)$
and $2\phi'$, $2\phi''$ (in units of $\pi$),
at the local maxima are also given. 
}
% % 3 LAYERS: Earth Mantle, Core, Mantle; 
% % Subject: Pem na ``grebne''
\vskip 0.4truecm
% \begin{center}
\begin{tabular}{|c||ccc||ccc||cc|}
\hline
\hline
h$^{\circ}$ & $\sin^22\theta$
& ${{\Delta m^2}\over E}$ 
      & $P_{\mu e}$
      & $\sin^22\theta'_m$  
      & $\sin^22\theta''_{m}$
      & $\sin^2(2\theta''_m - 4\theta'_m)$
      & $2\phi'/\pi$ 
      & $2\phi''/\pi$ \\
\hline

0. &  .034 &  7.201 &  1.000 & .111 & .614 &  1.000 & .923  & .949 \\
0. &  .024 &  7.194 &   .951 &  .082 & .549 &  .976  & .915 &    .855 \\
0. &  .020 &  7.191 &   .890 &  .069 & .509 &   .941 &  .912 & .810 \\
0. & .015  & 7.187  &  .760  & .051  &  .439 &   .854 &  .907 &  .746 \\
0. &  .010 &  7.183 &  .562  & .033  & .339  &  .695  &  .902 &  .677 \\
0. & .008  &  7.182 &   .500 & .028  & .307  &  .637  &  .901 &  .658 \\
0. & .007  & 7.181  &  .432  & .023  &  .271 &  .570  &  .900 &  .639 \\
0. & .005  & 7.180  &  .359  & .019  &  .231 &  .491  &  .898 &  .620 \\
0. & .004  & 7.179  &  .279  & .014  &  .185 &  .398  &  .897 &  .599 \\
0. & .003  & 7.178  &  .193  & .009  &  .132 &  .289  &  .896 &  .579 \\
0. & .001  & 7.177  &  .100  & .005  &  .071 &  .158  &  .895 &  .557 \\

\hline

13. & .039 & 7.024 &  1.000 &  .131   & .553 &  1.000 &  .931 &  .950 \\
13. & .030 & 7.018 &  .967  &  .102  & .498 &  .982 &  .923  & .878 \\
13. & .025 & 7.015 &  .916  &  .087   & .460  &  .950 &   .918 &  .837 \\
13. & .020  & 7.012  & .836  &  .071  & .415  & .895  &  .914  & .795 \\
13. &  .014 & 7.008 & .675   & .050   & .336  & .768  &   .908 &  .735 \\
13. & .009  & 7.005 &  .504  &  .034  & .256  & .613  &   .904 &  .686 \\
13. & .006  & 7.003 &   .362 &   .023 &  .189 &  .466 &   .901 &  .652 \\
13. & .005  & 7.002 &   .282 &   .017 &  .150 &  .374 &   .900 &  .634 \\
13. & .003  & 7.001 &   .195 &   .011 &  .106 &  .268 &   .898 &  .616 \\
13. & .002  & 7.000 &   .101 &   .006 &  .056 &  .145 &   .897 &  .597 \\

\hline

23. & .051 &  6.460 &  1.000 & .195 & .387 &  1.000  &  .910 &  .922 \\
23. &  .041 &  6.455 &   .976 &  .160 & .343 &  .985  &   .900 &  .876 \\
23.&  .031 &  6.450 &   .891 &  .123 & .288 &  .919  &   .889 &    .828 \\
23. & .024 &  6.447 &  .800  &  .100 & .248 &   .843 &    .883 &   .798 \\
23. & .020 &  6.445 &   .720 &  .084 & .217 &  .772  &    .878 &  .778 \\
23. & .016 &  6.443 &   .621 &  .068 & .184 &  .680  &   .874  &  .756 \\
23. & .010 &  6.440 &  .435  &  .043 & .125 &  .494  &   .867  &  .724 \\
23. & .008 &  6.439 &   .361 &  .035 & .103 &   .416 &   .865  &  .712 \\
23. & .006 &  6.438 &   .281 &  .026 & .080 &  .329  &   .863  &  .701 \\
23. & .004 &  6.437 &   .194 &  .017 & .055 &  .231  &   .860  &  .689 \\
23. & .002 &  6.436 &   .101 &  .009 & .029 &  .122  &   .858  &  .678 \\

\hline
\hline
\end{tabular}
% \end{center}
\end{table}

\vskip -3.0cm
\renewcommand{\arraystretch}{0.6}  % {\arraystretch}{0.5}
\begin{table} % \label{tab:tab2a}
\noindent \caption{The same as in Table 5 for 
the probability $P_{e2}$ ($\nu_e - \nu_{\mu (\tau)}$ mixing). 
}
% 3 Layers: Earth Mantle, Core, mantle
%  Subject: Pe2 na ``grebne''

\vskip 0.4truecm
% \begin{center}
\begin{tabular}{|c||ccc||ccc||cc|}
\hline
\hline
h$^{\circ}$ & $\sin^22\theta$
& ${{\Delta m^2}\over E}$ 
      & $P_{e2}$
      & $\sin^22\theta'_m$  
      & $\sin^22\theta''_{m}$
      & $\sin^2(2\theta''_m - 4\theta'_m + \theta)$
      & $2\phi'/\pi$ 
      & $2\phi''/\pi$ \\
\hline

13. & .044 & 7.100 & 1.000 & .145 & .611 & 1.000 & .953 & .974 \\
13. & .030 & 7.092 & .934 &  .102  & .536 & .965  &  .941 &  .859 \\
13. & .025 & 7.089 & .864 &   .084 &  .494 & .923 &  .936 &  .811 \\
13. & .019 & 7.086 &  .762 &   .067 &  .441 & .855 &  .931 & .761 \\
13. & .011 & 7.081 & .503  & .037   & .311 &  .640 &  .923 &  .669 \\
13. & .009 & 7.080 & .435  &  .031  & .275 &  .574 &  .922 &  .649 \\
13. & .007 & 7.079 & .361  &  .025  & .235 &  .495 &  .920 &   .628 \\
13. & .005 & 7.078 & .281  &  .019  & .188 &  .403 &  .919 &   .607 \\
13. & .004 & 7.077 & .195  &  .013  & .135 &  .292 &  .917 &   .585 \\
13  & .002 & 7.076 &  .101 &  .006  & .073 &  .160 &  .915 &   .563 \\

\hline

23. & .060 &  6.507 &  1.000 &  .223 & .434 & 1.000 & .931 &  .951 \\
23. & .024 & 6.492 & .724 & .097 & .256 & .782 &  .894 & .782 \\
23. & .012 & 6.487 & .439 &  .050 & .152 & .509 & .882 & .718 \\
23. & .010 & 6.486 & .365 &  .040 & .126 & .431 & .879 &  .704 \\
23. & .007 & 6.485 & .284 &  .030 & .098 & .343 & .877 & .690 \\
23. & .005 & 6.484 & .197 &  .020 & .068 & .242 & .874 & .676 \\
23. & .002 & 6.483 & .102 &  .010 & .036 & .129 & .872 & .662 \\

\hline
\hline
\end{tabular}
% \end{center}
\end{table}

\end{document}